# Revolutionizing Genomics with Reinforcement Learning Techniques

Mohsen Karami[#1], Khadijeh Jahanian[#2,3], Roohallah Alizadehsani[4], Iman Dehzangi[5,6], Juan M Gorriz[7], Yudong Zhang[8], Jia Wang[9], Farshid Hajati[10], Min Yang[11], Thantrira Porntaveetus[3, 12], Hamid Alinejad-Rokny*[13,14]

[1] Independent Researcher.

[2] Faculty of Engineering and IT, University of Technology Sydney, Sydney, Australia.

[3] Center of Excellence in Precision Medicine and Digital Health, Department of Physiology, Faculty of Dentistry, Chulalongkorn University, Bangkok, Thailand

[4] Institute for Intelligent Systems Research and Innovation (IISRI), Deakin University, Victoria, Australia.

[5] Department of Computer Science, Rutgers University, Camden, NJ, 08102, USA.

[6] Center for Computational and Integrative Biology, Rutgers University, Camden, NJ, 08102, USA

[7] Department of Signal Theory, Networking and Communications, Universidad de Granada, Spain.

[8] School of Computing and Mathematical Sciences, University of Leicester, Leicester, UK.

[9] Xi'an Jiaotong-Liverpool University, Suzhou, Jiangsu, China

[10] School of Science and Technology, Faculty of Science, Agriculture, Business and Law, University of New England, Armidale, NSW, 2350, Australia

[11] Shenzhen Institute of Advanced Technology, Chinese Academy of Sciences, Shenzhen, China.

[12] Clinic of General-, Special Care and Geriatric Dentistry, Center for Dental Medicine, University of Zurich, Zurich, Switzerland.

[13] UNSW BioMedical Machine Learning Lab (BML), School of Biomedical Engineering, UNSW SYDNEY, Sydney, NSW, 2052, Australia.

[14] Visiting Scholar (Collaborative Projects), Center of Excellence in Precision Medicine and Digital Health, Chulalongkorn University, Bangkok, Thailand

# Joint first author
* To whom correspondence should be addressed to H.A.R. h.alinejad@unsw.edu.au

## Abstract

In recent years, Reinforcement Learning (RL) has emerged as a powerful tool for solving a wide range of problems, including decision-making and genomics. The exponential growth of raw genomic data over the past two decades has exceeded the capacity of manual analysis, leading to a growing interest in automatic data analysis and processing. RL algorithms are capable of learning from experience with minimal human supervision, making them well-suited for genomic data analysis and interpretation. One of the key benefits of using RL is the reduced cost associated with collecting labeled training data, which is required for supervised learning. While there have been numerous studies examining the applications of Machine Learning (ML) in genomics, this survey focuses exclusively on the use of RL in various genomics research fields, including gene regulatory networks (GRNs), genome assembly, and sequence alignment. Additionally, we expand on RL's potential in cutting-edge areas such as single-cell genomics and its integration with large language models (LLMs), which hold promise for advancing medical and genomic applications. We present a comprehensive technical overview of existing studies on the application of RL in genomics, highlighting the strengths and limitations of these approaches. Special attention is given to the design and implementation of reward functions, which are critical for RL success, particularly in genomics tasks such as optimizing gene expression levels, aligning genomic sequences, and controlling gene regulatory networks. Furthermore, we discuss how RL can be integrated with other machine learning techniques to address complex challenges. We then discuss potential research directions that are worthy of future exploration, including the development of more sophisticated reward functions as RL heavily depends on the accuracy of the reward function, the integration of RL with other machine learning techniques, and the application of RL to new and emerging areas in genomics research. Finally, we present our findings and conclude by summarizing the current state of the field and the future outlook for RL in genomics.

## I. Introduction

Reinforcement Learning (RL) is a capability of the human brain that can be influenced by certain genes (1, 2) or may malfunction as a result of diseases such as schizophrenia (3, 4). RL was first introduced in research on adaptive control (5), and it was initially considered a subfield of adaptive control. Despite early challenges posed by computational limitations, RL is now recognized as one of the most effective learning methods across various domains, such as signal processing, image processing, and control theory. RL focuses on learning a policy for selecting actions in a given environment with the goal of maximizing the expected sum of discounted rewards, also known as utility (5). To achieve this, various methods have been proposed, including value and policy iteration methods, Monte Carlo methods, temporal difference methods, and policy search methods (5). RL is neither fully unsupervised, as it relies on reward signals, nor fully supervised, as it lacks labeled data, but it can adapt to changes through exploration (6). RL agents can learn to reach the desired objectives set by the environment without explicit instructions on how to do so (7).

The genome of a cell refers to its set of DNA, including its shape and organization, that contains all of its genes. The human genome contains three billion base pairs of DNA, which determine our basic anatomy and unique characteristics (8). Key genomic studies, such as gene regulatory networks and gene sequence alignment, play a crucial role in treating diseases such as cancer (9), developing safe and effective drugs, and advancing agriculture and plant science (10). Genomics, a subfield of biology, studies various aspects of genes, such as their function, structure, evolution over time, and their arrangement. With the development of various methods for obtaining genomic data (11-13), it is crucial to create fast, powerful, and adaptive tools for processing and analyzing them. Over the past decade, Machine Learning (ML), and specifically RL approaches, have been widely used in the analysis of genomic data (14, 15).

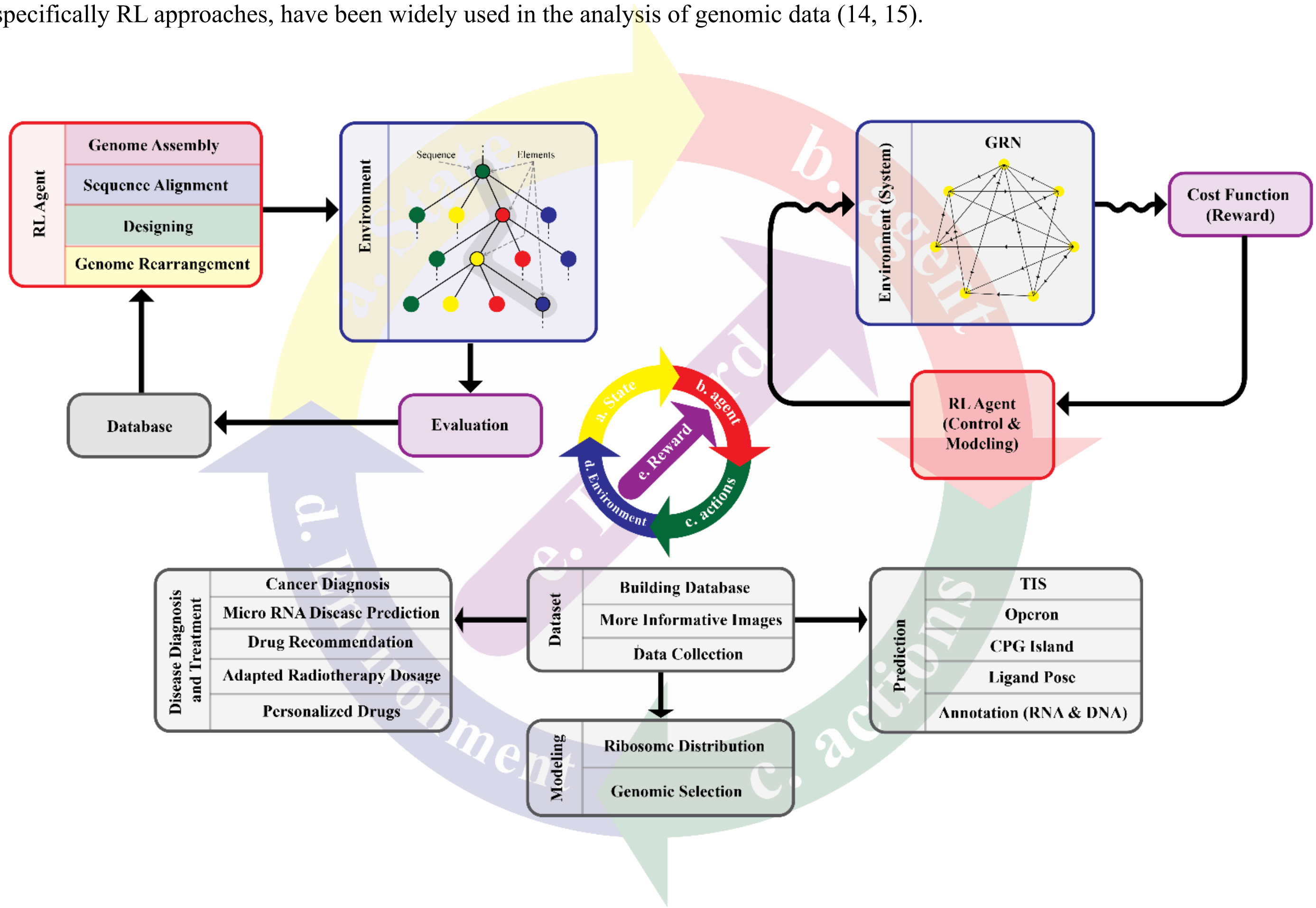

Figure *1* demonstrates the applications of RL in genomics-related problems that are reviewed in this paper (16).

The paper is structured as follows: In the next section, we outline our search strategy for paper selection. Section II provides a concise overview of RL. Moving forward, Section III delves into the application of RL in genomics, specifically focusing on fields like Gene Regulatory Networks (GRNs), Genome Assembly, and large language models.

We discuss the challenges within each area and review previous studies that have utilized RL to address these issues. Section IV outlines potential future research directions, while Section V presents the conclusion.

## *A. Reinforcement learning*

An RL agent aims to maximize expected sum of discounted rewards by appropriate action selection in successive time steps (17). Defining a reward function appropriately is critical to the RL success. The reward function must encode the necessary information needed to guide the agent toward achieving the environment objective. With each action performed by the agent, a reward is received, and the state of the environment is changed.

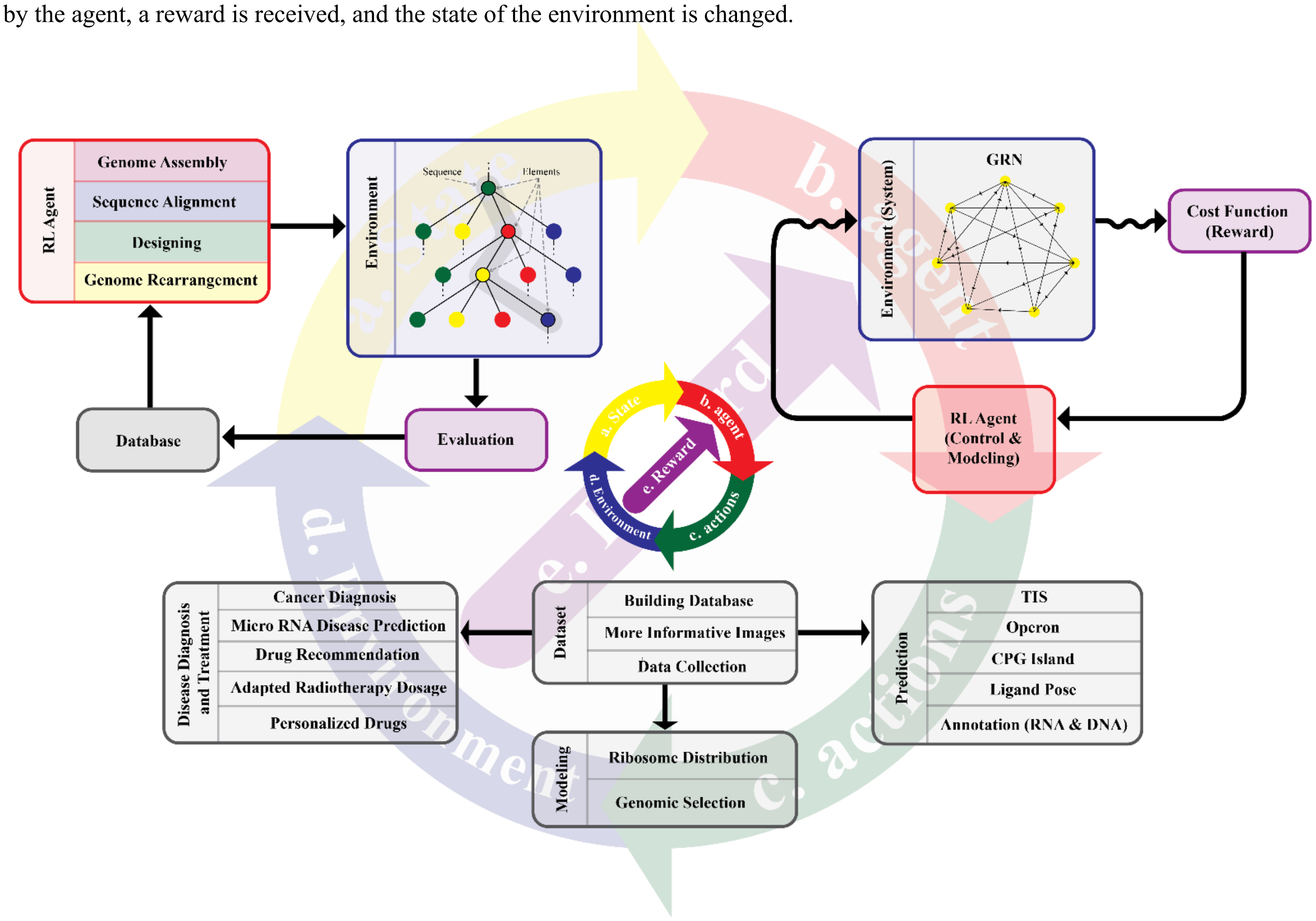

Figure *1.* The Role of Reinforcement Learning in Genomics: RL has primarily been employed for investigating gene regulatory networks. The second most widespread application of RL is in the field of sequencing, such as genome assembly and alignment. Despite being less frequently utilized for tasks such as modeling and prediction, RL has demonstrated superior performance compared to alternative methods.

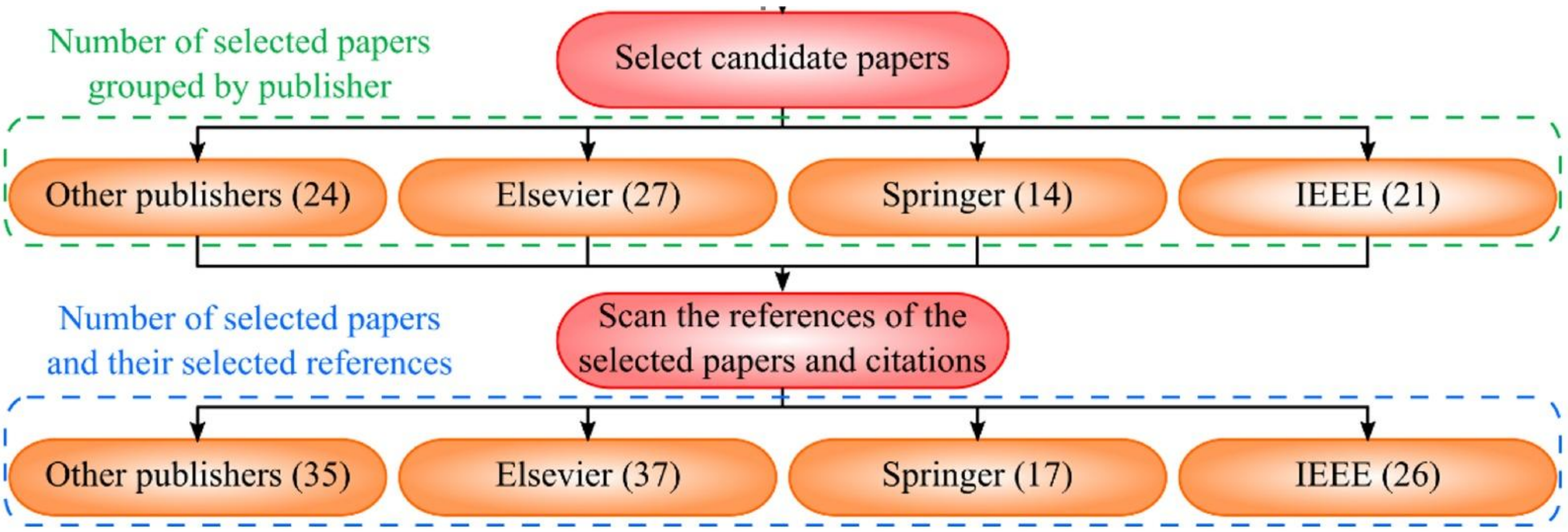

**Figure 2.** The flowchart of gathered papers in this survey.

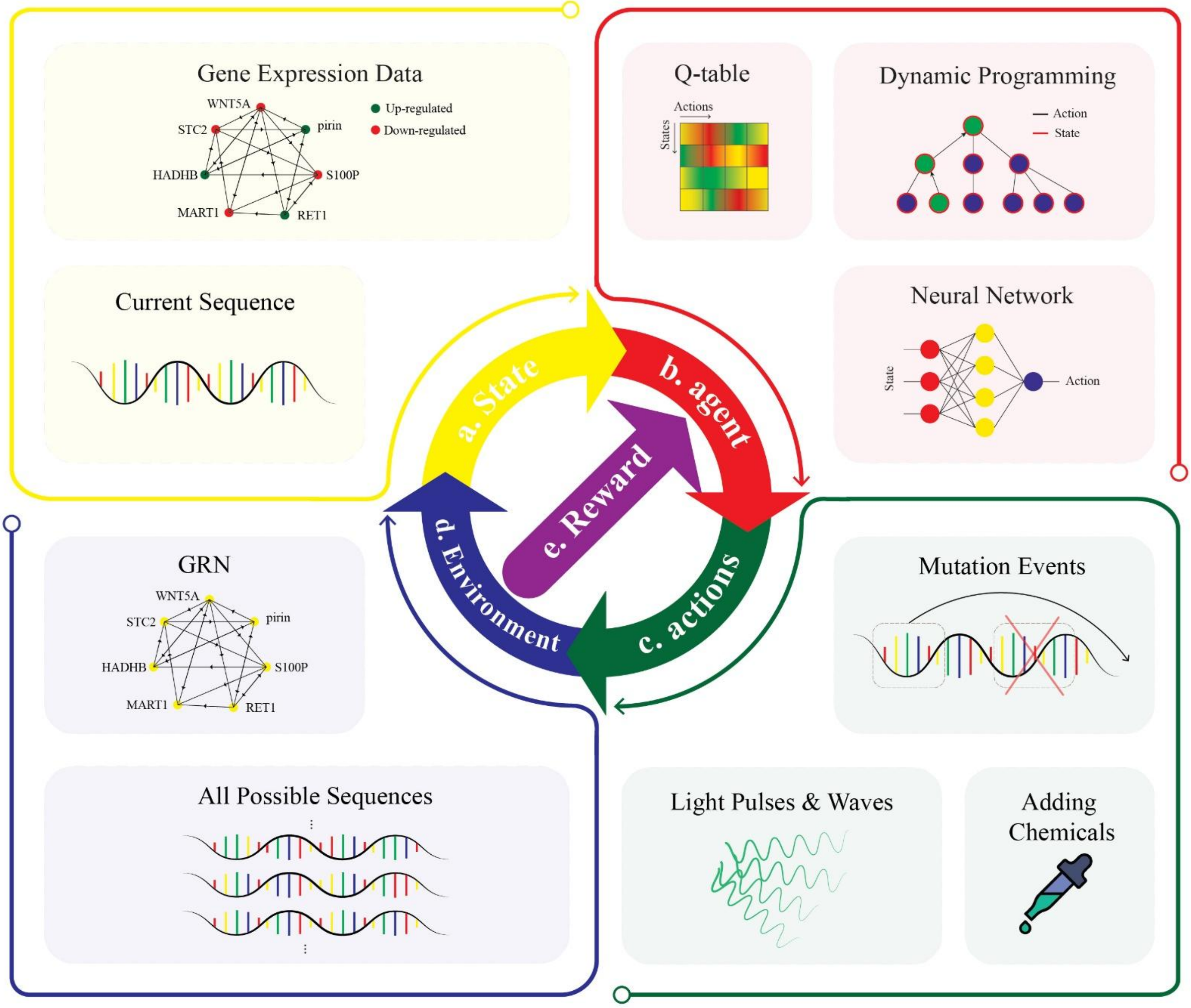

**Figure *3*** shows how an RL agent interacts with its environment. At each time step t, the agent uses its policy $\pi(a_t|s_t)$ to select action $a_t$ given state $s_t$. After action execution, the environment provides the agent with the next state $s_{t+1}$ and reward $r_{t+1}$. This process is repeated for a certain number of successive time steps until some termination condition is met. The set of transitions in the form $\{(s_t, a_t, r_{t+1}, s_{t+1}), t = 1, \ldots, T\}$ is called an episode with length T. The episode termination is triggered when the agent reaches the desired target or certain number of time steps is passed (17).

In RL, it is usually assumed that the future state $s_{t+1}$ is solely dependent on the current state $s_t$ and chosen action $a_t$. Such environment is said to be Markovian (17). The action selection of RL agents may be implemented as a policy or value function. **Policy function** $\boldsymbol{\pi(a_t|s_t)}$ determines the agent's behavior (action $a_t$) given observed state $s_t$. The policy function may be stochastic or deterministic. **The state value function** $\boldsymbol{V(s_t)}$ estimates the expected sum of discounted future rewards given that a particular policy is followed for action selection from the current state to the end of the episode.

In each RL problem, the environment behavior is governed by its time-transition model (dynamic model). After each action execution in the environment, the next state is determined by the dynamic model. RL algorithms that rely on the dynamic model during learning are called model-based. On the other hand, RL algorithms that solely rely on the set of observed transitions ($\{(s_t, a_t, r_{t+1}, s_{t+1}), t = 1, \ldots, T\}$) for agent training are called model-free (17).

## *B. Search strategy*

The study included papers gathered from Google Scholar, Scopus, and Pubmed search engines.

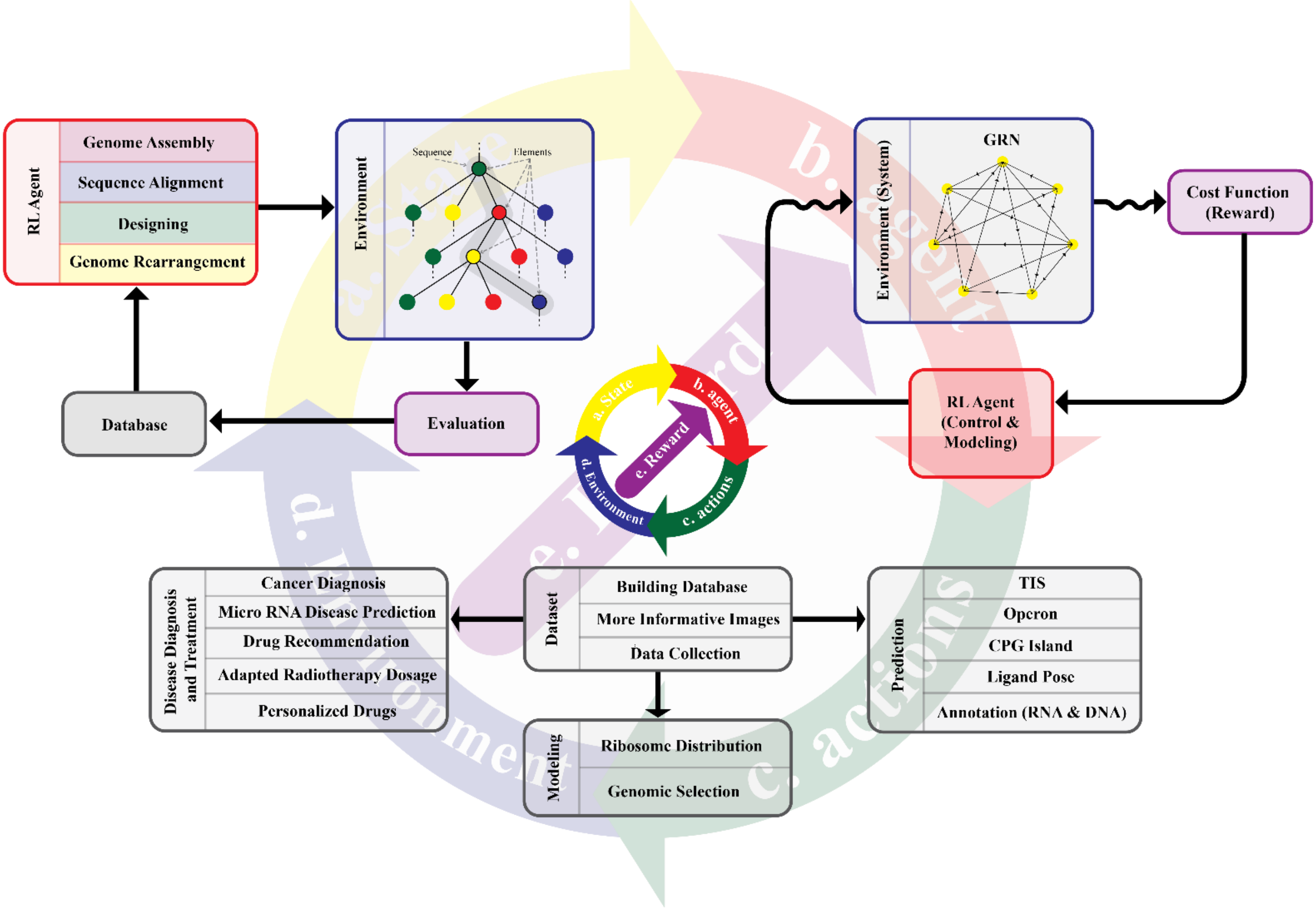

Figure ***1*.** The Role of Reinforcement Learning in Genomics: RL has primarily been employed for investigating gene regulatory networks. The second most widespread application of RL is in the field of sequencing, such as genome assembly and alignment. Despite being less frequently utilized for tasks such as modeling and prediction, RL has demonstrated superior performance compared to alternative methods.

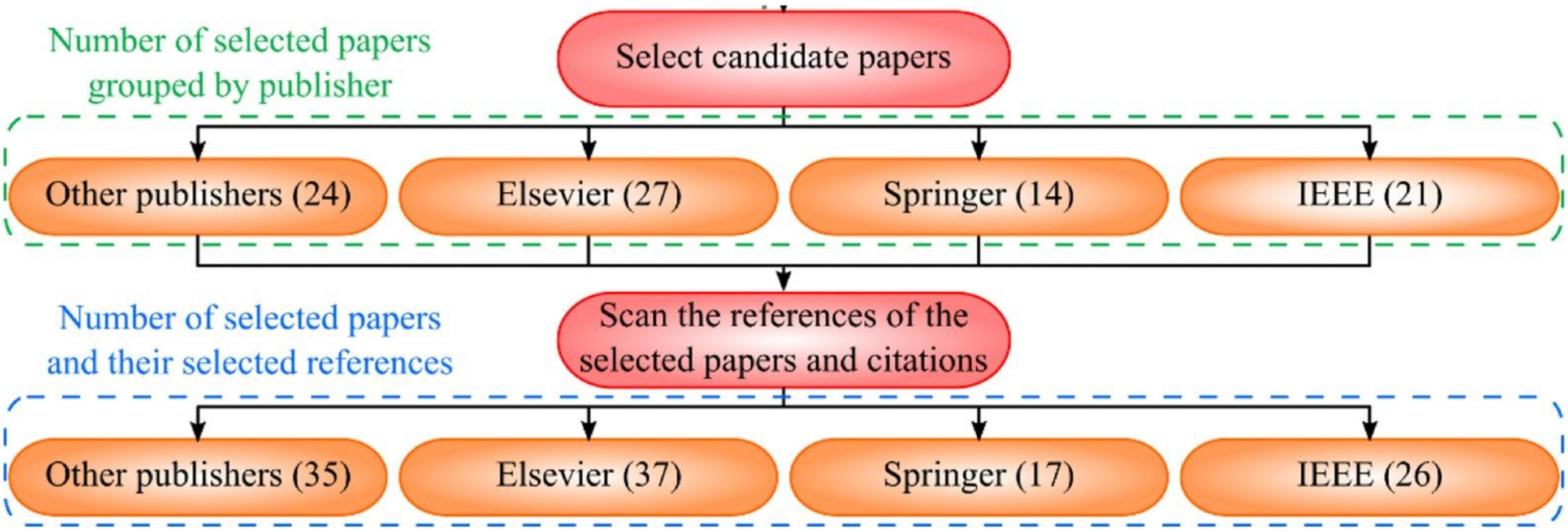

**Figure *2*** s illustrates the number of reviewed papers, obtained through a two-phase search strategy. The top row represents the papers selected in the initial phase, while the bottom row includes the papers and their selected references. Papers were grouped based on their publisher names. The list of reviewed papers was compiled by four authors, with inclusion based on approval by at least three authors. Selection criteria included publishers, citation count, and relevance to RL

applications in areas such as gene regulatory networks, sequence alignment, genome assembly, rearrangement, feature prediction, and dataset preparation/collection efficiency.

### C. Value-based methods

The state value function, $V(s_t)$, depends solely on the input state, $s_t$. However, it is also possible to consider the action taken at a certain time step, $a_t$, when estimating the value, leading to the action-value function, $Q(s_t, a_t)$, also known as the Q-function. The Q-function estimates the cumulative discounted reward that can be obtained in state $s_t$ by selecting action $a_t$ (18). By solving the Bellman equation and calculating the value function for different policies, the best policy that results in the highest value function can be determined. Methods for analyzing genomics data based on value functions can be grouped into two categories: dynamic programming and temporal difference. In dynamic programming, the defined cost function is maximized by moving backward in time from the best final state. This method has two main approaches: policy iteration and value iteration. Both approaches have two steps: policy evaluation and improvement. In the policy iteration approach, the Bellman equation is iteratively solved in the evaluation step, while in value iteration, it is only solved once to estimate the value estimation (19). On the other hand, temporal difference does not require a model and estimates the value function based on the observed reward. If applied to the action value function, it is referred to as the SARSA (State-Action-Reward-State-Action) method, which is an on-policy approach. The off-policy variant is called Q-learning (18), in which the policy being learned is different from the policy used to select actions (20).

### D. Policy search

In this method, the action selection is done using a policy function $\pi_\theta(a_t|s_t)$ with learnable parameters vector $\theta$. The optimal policy cannot be determined by applying Bellman equation recursively (21). Some of well-known policy search methods are policy gradient (which optimizes a parametrized policy function, regarding the reward), expectation maximization, and path integral (21).

**Actor-critic:** This architecture consists of an action selection policy (actor) and a value function (Critic) (22). As the name implies, the critic judges the actor performance in terms of expected discounted reward provided that the actor is used for action selection until the end of the current episode. The critic does not take part in action selection, but it is paramount in actor parameters learning. The actor is updated such that the score it receives from the critic is maximized (22).

**Deep RL:** DRL employs Deep Neural Networks to implement agent policy, value function, and transition model of the environment. Thanks to the representation power of DL, high dimensional RL problems e.g. genome dynamic systems, that were previously deemed unsolvable can now be undertaken efficiently (23).

**Inverse RL:** In some cases, such as social behaviors, analytical calculation of the cost function and reward value may be difficult or impossible. In such scenarios, IRL has proved to be useful since it can infer the reward function based on a given policy or expert behavior (24).

### E. RL challenges

There are several challenges when using RL methods in optimization problems, which are discussed below:

**Exploration vs. exploitation:** To implement successful learning, an agent must strike a balance between exploration and exploitation. Exploration is necessary for the agent to discover new regions of its environment, while exploitation allows the agent to make use of what it has learned during the training phase in order to maximize the expected sum of discounted rewards (17). Maintaining this balance is one of the crucial challenges of RL. Exploration is typically achieved through the use of noise-based perturbations, such as Gaussian noise, the Ornstein-Uhlenbeck process (25), or parameter space noise (26).

**State-action space dimensionality:** As the number of states and actions increases, so does the dimensionality of the problem, leading to increased computational complexity (18). While deep reinforcement learning models have demonstrated the ability to handle high-dimensional problems, their training requirements, such as the use of graphical processing units, can be prohibitively expensive (27).

**Preparation of training data:** In RL, the training samples are collected through interactions with the environment, which can be time-consuming and costly. Furthermore, real-world data is often noisy and may contain errors due to sensor failures, which can undermine stable learning (17).

**Unknown environment model:** Access to an accurate environment model can simplify the RL simulation problem, but such models are often unavailable in practice (28).

**Reward function definition:** Designing appropriate reward functions that reflect the goals of the problem is crucial for successful RL. This can be particularly challenging in complex scenarios, and alternative approaches, such as sparse reward functions (which are given only when the agent reaches its specific goal at a terminal state) or Hindsight Experience Replay (which reuses agent memories stored in a buffer to improve stability) (29), have been proposed to address the issue of sparse rewards. In particular, sparse reward functions pose a challenge for classic RL agents (30), as they provide no indication of how close the agent is to reaching the goal, hindering effective learning.

## II. Application of Reinforcement Learning in Genomics

Previous studies have demonstrated the critical role of machine learning techniques across various domains of biomedical research, including genomics data analysis, medical image processing, and medical signal processing. In genomics data analysis, machine learning has been instrumental in identifying gene-disease associations (31-36), identifying copy number associations (37, 38), and interpreting non-coding regions of the genome (39-44). In medical image processing, machine learning methods have enabled advancements in tasks such as tumor segmentation, early disease detection, and classification of imaging biomarkers from modalities like MRI, CT, and histopathology images (45-47). Meanwhile, in medical signal processing, these techniques have shown significant impact in analyzing physiological signals such as ECG, EEG, and EMG, contributing to improved diagnosis and monitoring of cardiac, neurological, and heart conditions (48-54). Collectively, these applications underscore the transformative potential of machine learning in modern biomedical research. RL is a widely used and effective tool in various fields, including genomics. It has been utilized in creating adaptive, optimal, and autonomous intelligence control solutions. In the context of genomics, RL has been applied to a range of problems and this section aims to provide an overview of its applications. Our literature searches were focused on human and mice English language papers available in the PubMed, Scopus, and Web of Science. The search terms included "Reinforcement Learning", "Genomics", "Medical Data Analysis", "Gene Regulatory Networks", "Genome Assembly", "Sequence Alignment".

### *A. Gene regulatory networks*

A gene regulatory network (GRN) is considered a set or subset of genes interacting with one another to control and perform specific cell functions (55). Genes are often considered as the baseline of a biological entity, specifically DNA strand subsequences containing consecutive protein-coding regions. Gene regulatory networks are essential in developing, differentiating, and responding to environmental cues (56, 57).

As

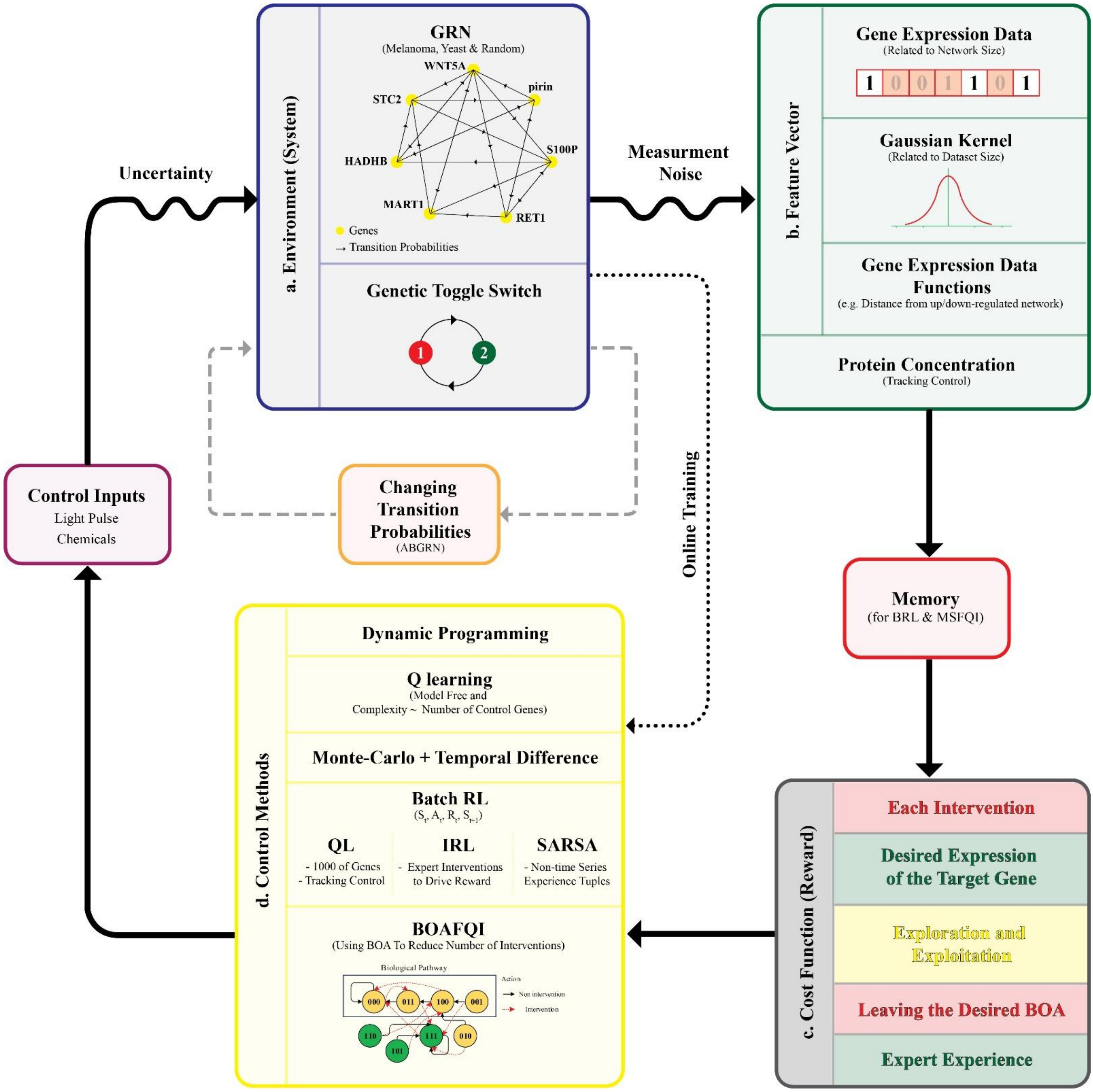

**Figure** *4* shows RL has been used in modeling and controlling GRNs and Probabilistic Boolean networks (PBNs) which is a subclass of Markovian GRNs (58). In a PBN, each gene value is either 0 or 1 corresponding to the Up and Down-regulated states. In part a of

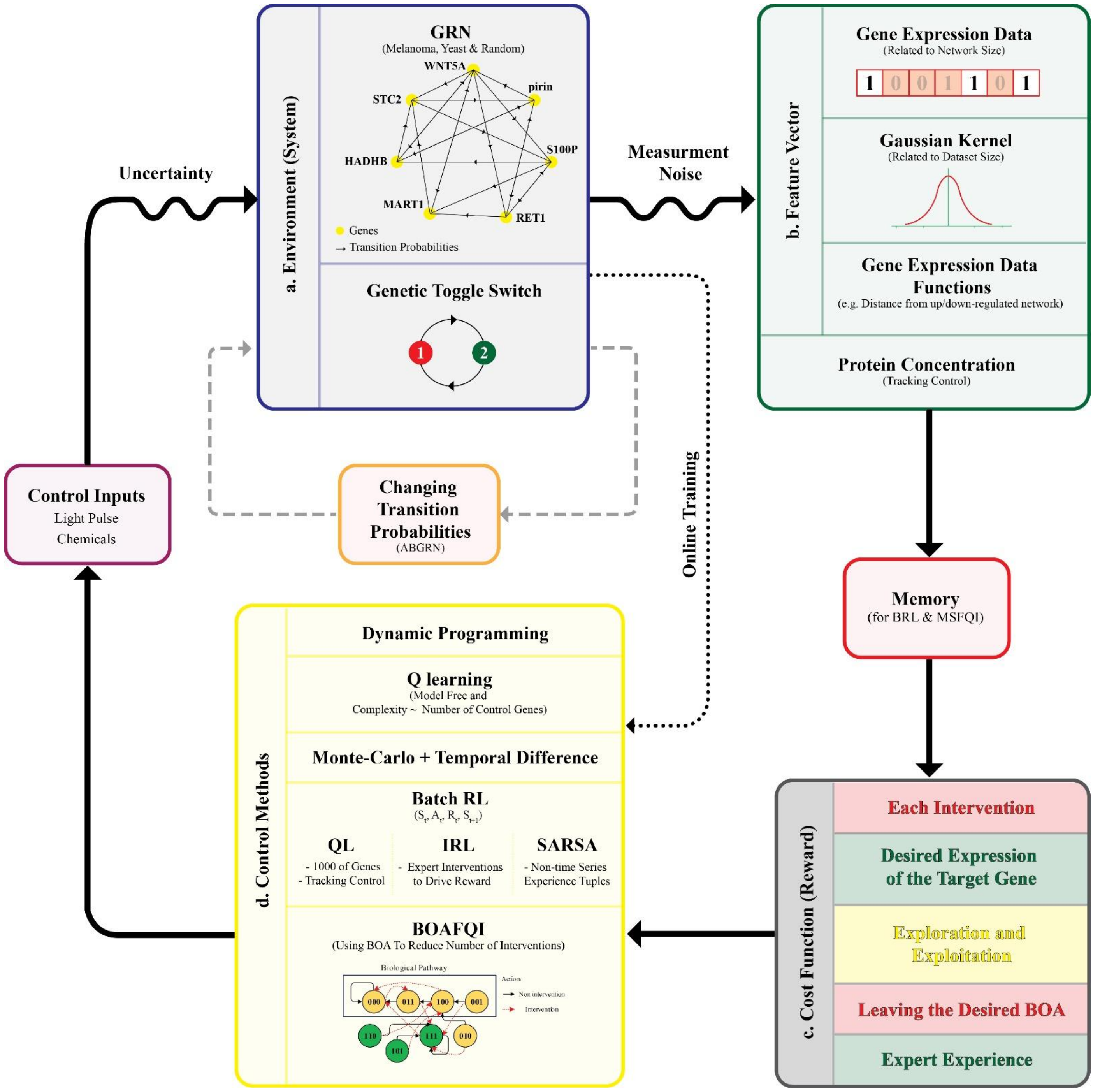

**Figure *4***, a typical 7-gene PBN (59) is shown which is commonly used in many papers.

In PBNs, state transition occurs based on the network current state and fixed transition probabilities. However, in order to move toward our desired state, the network behavior can be controlled by altering the transition probabilities based on some external input. The optimal control input can be determined by RL. For instance, dynamic programming has been employed to determine optimal intervention strategies for each state via cost function minimization (59).

According to Bittner et al. (32), *WNT5A* is a very important gene related to Melanoma. In this study, Dynamic programming is used with a fixed penalty for the final state indicating that the best final state is when the *WNT5A* state is 0 or down regulated. The result shows that the optimal control increases the probability of the desired state from 0 to 0.673. However, dynamic programming requires transition probabilities (model) and its computational complexity scales as the number of genes increases. To keep the computational complexity manageable, a model-free approximate stochastic control based on Q-learning has been proposed (60) where the number of genes in the network depends on the number of binary control inputs. To deal with the slow convergence of the method, the network and model proposed in (59) were used to drive the necessary data and assigned a penalty for each intervention.

Apart from the higher computational complexity of model-based RL methods, learning environment dynamic model is often challenging and susceptible to errors which is why model-free RL methods are still popular. Sirin et al. proposed a model-free RL approach called Fitted Q Iteration (FQI) to find the optimal policy for controlling GRNs. To this end, the gene expression data was considered as transitions collected from the environment in the RL setting. The Q-function used in this approach consists of features (gene expression) and parameter vectors that are estimated directly from collected experiences via the least square optimization(61).

The performance of LSFQI method was improved by using Gaussian kernel function as feature vector (62) but at the cost of higher time complexity and memory usage. The aforementioned overhead was due to the larger size of the employed feature vector. LSFQI has also been used to control GRN for tracking a periodic reference signal in a limited time step (63). Protein concentration was used as the control variable, and light pulses as control signals in a 6-gene generalized repressilator network. It was reported that in the reference signals with short periods, performance drops and non-control state variables oscillate dramatically. This method can only be used for oscillatory behaviors and has significant limitations on the reference signal.

Previous methods proposed for GRN control based on batch RL used a single experiment due to absence of online training. However, relying on a single experiment may end up in training with poor quality. The issue was addressed by utilizing a hybrid method combining online and offline training (63). The hybrid method achieved successful control in uncertain and stochastic dynamics.

Learning from interactions with partially observable environment is desirable which has inspired Sirin et al. (64) to assume partial observability for GRN and use Batch RL method called least squares fitted TD(λ) iteration (LSFTDI) to combine Monte-Carlo (MC) and temporal difference (TD). To this end, three different feature vectors for Q-function estimation based on Melanoma and yeast datasets have been studied. The gene expression vector was reported as the best feature vector. Experimenting with the model on large PBNs revealed that as the number of genes increases, the performance gets closer to the uncontrolled system, but the convergence time increases as well.

Unlike LSFTDI which requires samples in the form of time-series, Nishida et al. (65) proposed a method to break the dependence on time-series data leading to more training samples and the possibility to use data from different experiments. FQI SARSA was used to control the partially observable GRNs studied in (64). The major drawback of this work was that the validation model has the same number of genes as the training dataset, which is inconsistent with partial observability assumption. Uncertainty is entwined with partial observability. Handling uncertainty is crucial for successful learning under partial observability. Imani and Braga-Neto (66) assumed uncertain inputs and measurements in the 7-gene Melanoma GRN and modeled uncertainty using a Gaussian expression. Sparse approximation of Gaussian process SARSA (SGP-SARSA) was done to estimate Q-values for a set of state-action pairs spanning the whole space with reduced complexity. It was shown that a Gaussian process model can address the exploration/exploitation trade-off. However, this method is model-based, so it cannot be used in real networks. Imani and Braga-Neto extended their work (67) by proposing a new model in which in addition to the assumptions in their previous work (66), the reward (cost) function was assumed to be unknown. The estimation of the reward function was done using Bayesian Inverse RL (BIRL) method based on noisy trajectories data collected during interventions of an expert (e.g., a physician or biologist). This method uses the Markov chain Monte Carlo (MCMC) technique for reward estimation and combines it with Q-learning for faster convergence. Boolean Kalman filter (V_BKF) was used for state feedback and reward function estimation for network control. Results show successful estimation and control, but the time and computational complexity were not analyzed.

On one hand, solutions that disrupt a biological pathway are costly and may cause other diseases due to ignoring the biological side of the problem. On the other hand, the state of a GRN can be changed by perturbation or through its biological pathway. After completing a biological pathway, in the absence of perturbation, most GRNs cycle in the same set of states called attractors. The states that guide a GRN toward attractor states belong to basin of attraction (BOA). At the bottom of part d of

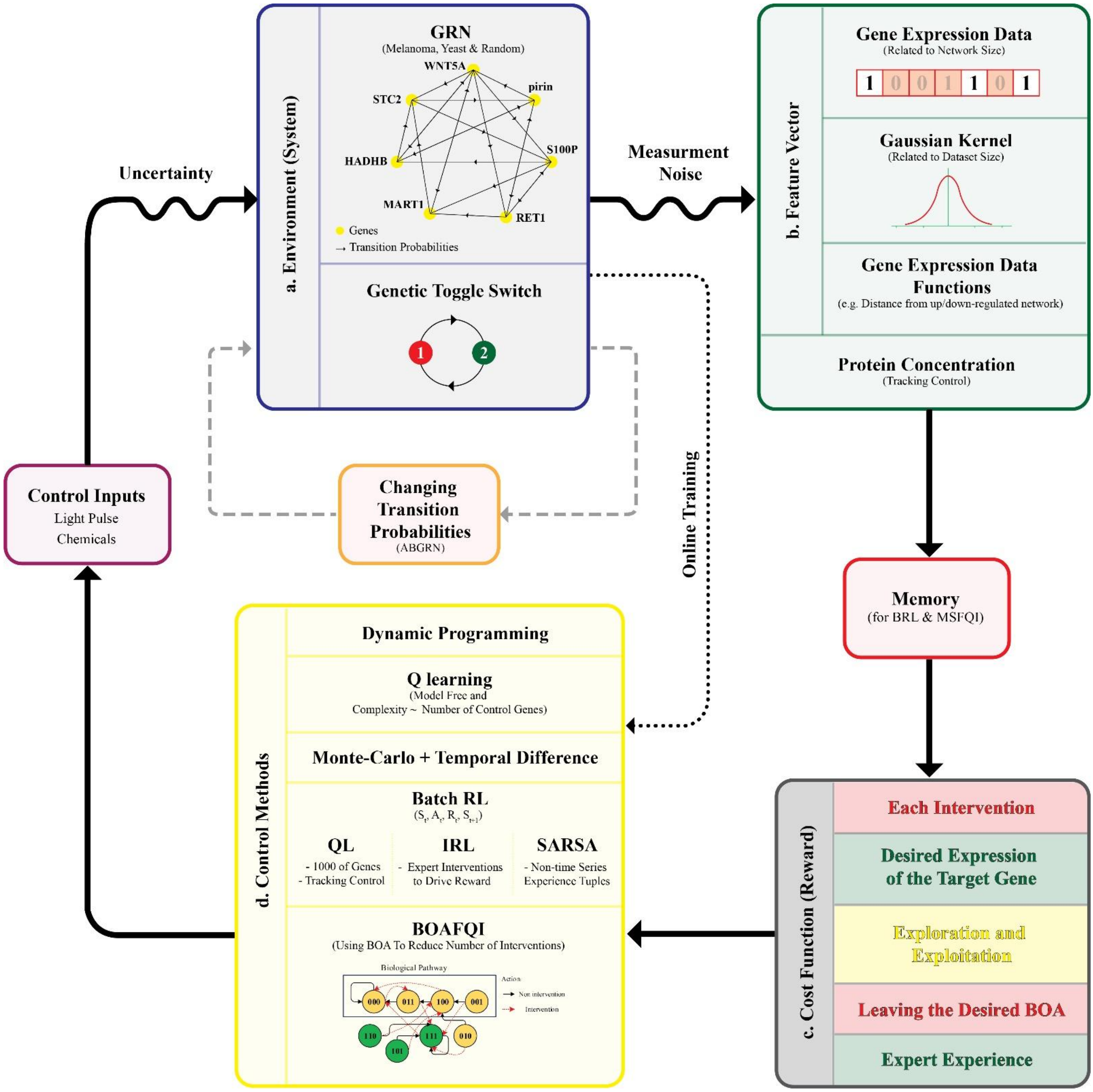

**Figure** *4*, a simple biological pathway in a 3-gene network is shown. As can be seen, the pathway is a graph-like structure similar to a finite state machine with attractor nodes “000” and “111” (68). As can be seen, the GRN can be guided to a desirable BOA after which GRN does the rest. Inspired by the given argument, a control method called the basin of attraction FQI SARSA (BOAFQI-SARSA) has been proposed (68).

The major challenge of guiding GRNs to healthy phenotypes is determination of the appropriate types of interventions and the right time of applying them. The guidance of GRNs is realized via gene activity profiles called BOAs. Determining health status of current phenotype is challenging due to partial observability of the genes. To address this challenge, the intervention strategies were defined by proposing a new batch RL method called mSFQI (69). The intervention strategies are calculated based on previous observed experiences.

From the literature, it can be concluded that the best methods are model-free and derive control signals from data directly because developing a model is not practical in real cases due to high number of genes. To sum up, RL methods show satisfying performance thanks to their adaptive, locally optimal, and model-free characteristics. More details on the reviewed literature can be seen in Tables caption

**Table** *1*.

## *B. Sequencing and rearrangement*

Developing new drugs and treatments with molecular sequences using an experimental approach is a costly and time-consuming process. There is a demand for fast and effective computational methods for designing molecular sequences using learning techniques such as RL (70, 71). Some previous studies used patient's genomic features to optimize the drug structure (molecule sequences). However, this section is devoted to existing literature on genomic sequences. Examples of these sequences are amino acids, DNA with action set {A, T, C, G}, and antimicrobial peptides (AMPs) with relatively short (8-75 amino acids) protein sequences. The motivation behind designing these sequences is reward function optimization which is complicated and time-consuming.

To tackle this problem, DyNa-PPO (proximal policy optimization) (72) was proposed that trains the policy offline and online with more data efficiency. This method estimates the reward function using a small experimental dataset during the first training phase which is online and model-free. The reward function is then used for offline and model-based policy training. For performance measurement, the cumulative reward and diversity of results were considered based on which DyNa-PPO proved to be superior to other evaluated methods except PPO.

One of the motivations for using RL in medical applications is its cost effectiveness. As an example, while biomolecular design has gained attention, its application is limited by the increasing cost of developing new drugs and treatments. RL approaches can be used to remedy the cost of working with biomolecular design.

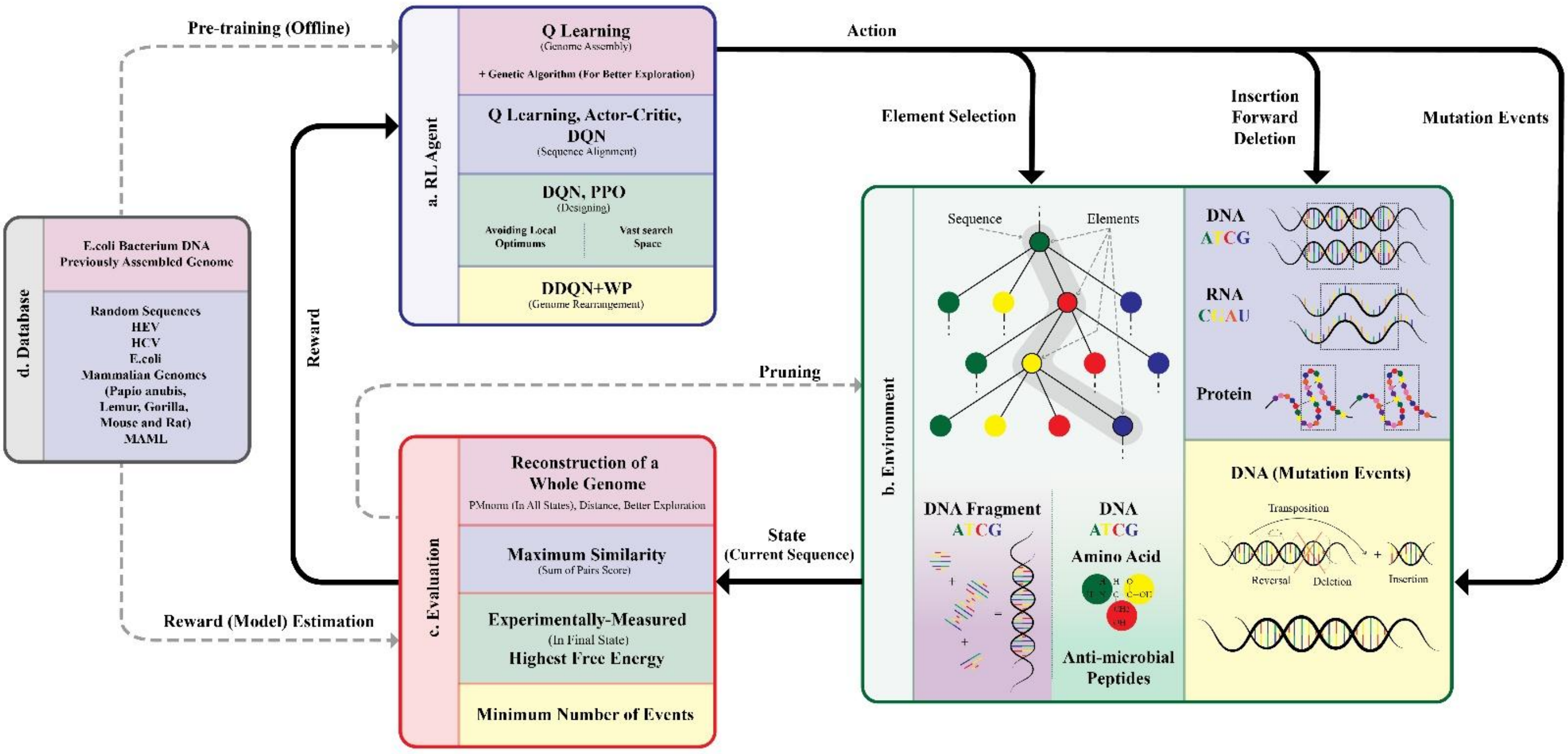

**Figure *5*** demonstrates the steps taken to solve sequencing and rearrangement problems based on RL. To this end, two RL methods and one greedy hill climbing search have been evaluated (73). The reward function was defined based on free energy calculation for a given RNA sequence (74). Among the evaluated RL methods, Deep Q-Networks (DQN) managed to outperform PPO and the greedy hill climbing search algorithm.

### *B.1. Sequence alignment*

In sequence alignment, DNA, RNA, or protein sequences are arranged to extract similar regions which may result from functional, structural, or evolutionary relationships between the sequences (75). For instance, a DRL method called DQNalign (76) has been proposed for local alignment that repeats on sub-sequence pairs (a moving window) over the two sequences for memory and time complexity reduction. The actions in local alignment are forward, insertion, and deletion. Upon action execution, the agent gets its reward for a gap, match, and mismatch (77).

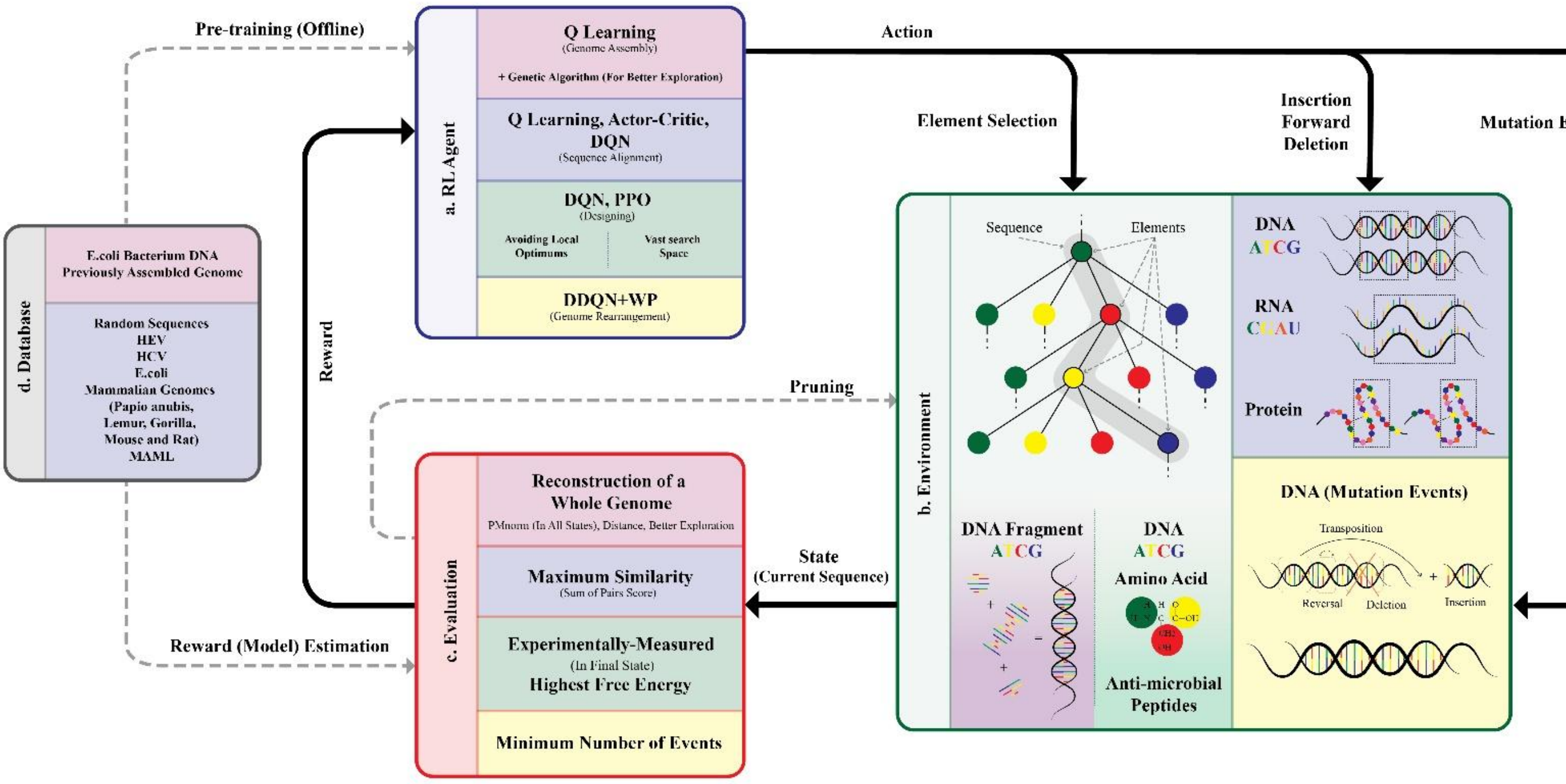

**Figure 5.** Interaction and training loop of RL in sequencing and rearrangement: (a) The RL agent selects actions. (b) The selected actions are executed in the environment, whose state is represented by a sequence, such as DNA or molecules. (c) The cost function provides the agent with reward values, based on which the agent optimizes the structure. (d) Available datasets are used for pre-training and reward estimation.

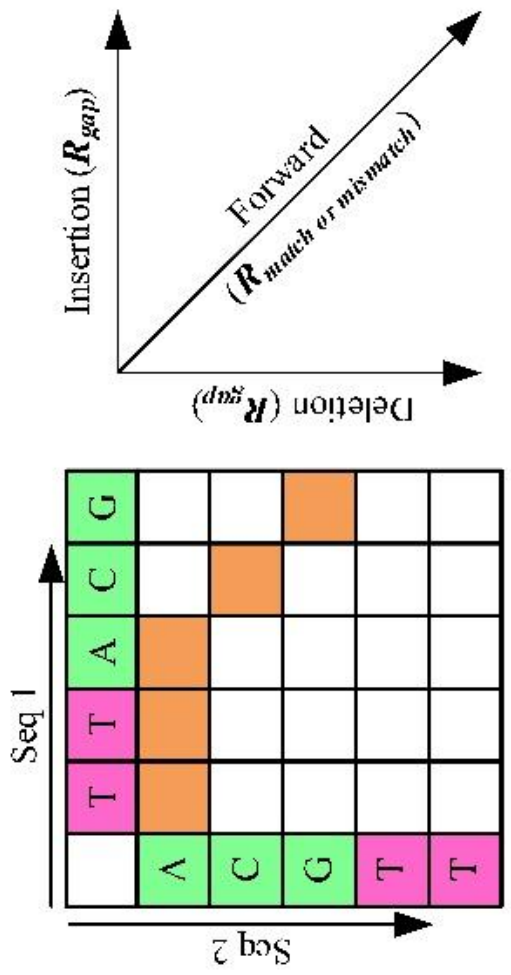

**Figure 6** shows the state-action space and the reward function in this problem (76). For finding the best window size, random sequences with their permutations were used. The performance of DQNalign method has been evaluated on Hepatitis E Virus (HEV) and E. coli datasets. Results show that the method outperformed conventional methods in sequences with similarities below 80%. However, an appropriate starting point is crucial for good performance due to single-direction sequencing. Moreover, the use of simulated training dataset may cause learning undesirable patterns.

Song and Cho (78) followed the same method but employed double DQN (DDQN) and X-drop algorithms to improve termination conditions and prevent unnecessary computations. X-drop terminates the alignment when the score falls below the highest score obtained by the agent considering gap X. Each training episode uses two random sequences for training within inner loop. The two random sequences are produced with model-agnostic meta-learning (MAML) and their mutations are used to update Q-network for the next episode. Larger window sizes and X values improve performance but increase time complexity linearly. In case of taking ε-greedy approach, high fluctuations may occur which leads to false terminations.

Alignment of three or more biological sequences with similar lengths for maximizing their similarity regions is called multiple sequence alignment (MSA) (79). These sequences are assumed to be connected via evolutionary relationships. MSA is considered an NP-hard problem which has been tackled utilizing Q-learning (80). The sum of pairs (SP) score (81) between all sequence pairs was considered as the reward function. Classic methods such as Mummer or Clustal were selected for pairwise alignment and experiments were carried out on real-life DNA datasets such as Hepatitis C Virus, Papio Anubis, Lemur, gorilla, mouse, and rat. The proposed approach was superior in 95% of the conducted experiments. The success of Q-learning inspired researchers to tackle the same MSA problem as in (80) by relying on the famous actor-critic architecture of RL (82). The actor network is updated based on policy gradient. The critic is implemented as a long, short-term memory (LSTM) and it is updated based on Q-learning. The time complexity of this method is not exponential because of the LSTM layers. The method scales with stat-action space leading to acceptable performance and exploration ability.

***B.2. Genome assembly***

Genome assembly (GA) is a computational representation of genome sequencing that involves analyzing and combining several small fragments of a large genome (reads). GA's goal is to reconstruct the complete human genome. The quality of the genome assembly is related to the various positive impacts it can generate for society, like improving the comprehension of living beings or developing effective treatments for diseases (83). Genome assembly is mostly done by comparative strategy and de novo strategy, which does not need a previously assembled genome as a reference (there are a few available), but it is a highly complex method. De novo strategy is mainly based on heuristics and graphs (Overlap-Layout-Consensus and De Bruijn graph). Each read considers a node, and assembly task is to find the optimal path between these nodes, similar to the traveling salesman problem (43).

Chemical limitation prevents reading the whole genome at once. That is why DNA fragment assembly (DFA) involves analyzing and combining several small fragments (reads) of DNA (84). DFA contains three phases that are overlap (finding the longest match), layout (ordering), and consensus (determining Original DNA).

Bocicor et al. (85) utilized Q-learning to tackle DFA problem. To this end, the sum of the overlapping scores (86) was used as the reward function. The optimal policy to produce the original DNA for 25 base pairs (bp) fragments of E.coli bacterium DNA was obtained by following greedy strategy based on estimated Q-values. Convergence to the optimal Q-values was achieved after $4 \times 10e6$ episodes and outperformed heuristic methods like the genetic algorithm and clustering. However, high computational complexity makes this method impractical for large datasets of real-world problems.

Application of Q-learning for DFA was further investigated on a more extensive dataset (87). Another metric in reward function also compared the assembler's output to original micro-genome (distance). The training and evaluation phases were conducted three times on each sample with five different numbers of episodes (1-5 million). However, this approach does not scale with the number of reads due to exploding dimensions of the Q-table. To overcome this drawback three strategies were considered (88). As the first strategy normalized PM or PMnorm was included in the reward function which considers the relative order of reads and gets different values depending on the state. The second strategy is action elimination by pruning which deletes actions that lead to fewer non-terminal rewards (88). The resulting method was aimed to improve state space exploration and sample efficiency. The third strategy combines RL with a genetic algorithm for better exploration. It uses the result permutation of each episode as an individual for the Genetic Algorithm's initial population. The best individual is then used for running the next episode. Additionally, larger micro-genomes (4kpb in the most significant case) have been used to reach approximately four times better performance according to distance/reward-based metrics in a much shorter time (88).

***B.3. Genome rearrangement***

Genome rearrangements are mutation events (non-conservative, like deletions and insertions, or conservative, like reversals and transpositions) that affect large fragments of the DNA and have important role in the disease progression (89, 90). Several machine learning-based toolsets have been developed to identify genome rearrangements however, RL methods still need to be investigated in this era more broadly. One of the genome rearrangements is Reversal in which a DNA fragment is reversed. Sorting by Reversals (SbR) calculates the minimum number of events (rearrangement distance) to transform one genome into another by reversal (91). To address SbR problem for a set of fragments, different RL algorithms like TD-Lambda and Double Deep Q-Network (DDQN) have been used (92). The RL agent is penalized with a reward value of -1 for taking each step. This way the agent is forced to find the minimum number of events. The action space (all reversals events) and state space (all permutations) of the problem is huge, so estimating the Q-table requires substantial training time and memory space. Although Kececioglu and Sankoff's greedy algorithm (93) was used for pre-training (92), the TD-lambda method did not converge for $n>8$. However, DDQN outperformed conventional methods in about one-quarter of cases.

In Sorting by Reversals and Transpositions (SbRT) problem, transposition occurs when two fragments exchange places. DDQN promising performance inspired its application alongside Wolpertinger Policy (WP) for tackling SbRT problem *(94)*. WP is based on actor-critic framework combined with the k-Nearest-Neighbor algorithm to deal with continuous action space. Results show that the WP delivers less distance compared to DDQN. However, WP has limited performance in high-dimensional state-action spaces (n>15).

Sequence alignment problems have an inherent complex nature, but RL has the potential to handle the aforementioned complexity. The structure of the genome assembly problem is very similar to sequence alignment, so it comes as no surprise that RL performs quite well in genome assembly. A summary of RL methods used for sequencing and rearrangement is presented in Tables caption

**Table 1.** A comparison between the reviewed techniques in this study based on performance, network size and their main contribution.

| | Network type | Method | Performance (%) | Key points |
|---|---|---|---|---|
| Control | 7-gene WNT5A | Dynamic programming | 67 | One of the first studies that used RL. |
| | 10-gene Melanoma-related | Q-learning | 67 | Model-free and linear complexity |
| | 8067-gene Large scale Melanoma | Bach RL (FQI) | 80 | Large networks |
| | 7-gene Partially observable melanoma | LSFTDI | 90 | Partial observability |
| | 6-gene generalized repressilator | FQI | Tracking was done | Tracking control |
| | 2-gene genetic toggle switch | FQI | Regulation was made | Online and Offline training with no prior knowledge |
| | 7-gene melanoma and yeast | FQI SARSA | 53 | No need for time series data |
| | 7-gene melanoma | SGP-SARSA | 90 | Uncertain inputs and measurements, model-based |
| | 11-gene yeast cell cycle and 10-gene Metastatic Melanoma | BOAFQI-SARSA | 99 | Partial observability, used BOA |
| | -gene Melanoma and a p53-MDM2 negative feedback | BIRL+V_BKF+ Q-learning | 90 | Partial observability, unknown reward |
| | 7 and 25-gene melanoma | SQFT | 95 | Different feature vectors, more exploration |
| | 1000 to 7000-gene synthetic BNps | mSFQI | 87 | Used BOA, partial observability |
| Modelling | coupled oscillations in the cell cycle & cAMP receptor | - | 95 (accuracy) | Used ABGRN |

**Table** *2*.

## C. *Other applications of reinforcement learning in genomics*

In addition to sequence alignment and rearrangement, RL has been utilized in other topics in genomics such as predicting, modeling, diagnosing, and treating diseases. The related literature is summarized in Figure 7.

### *C.1. Feature prediction*

The protein-encoding process contains two stages which are transcription and translation. During transcription, an mRNA molecule is synthesized using DNA sequence. The translation process involves synthesizing a protein molecule using mRNA, which contains three stages namely initiation, elongation, and termination. The translation process begins in Translation Initiation Sites (TIS) (95). In (96), a multi-agent system for predicting TISs called MAS-TIS was presented which used three different agents to improve the robustness and performance of overall TISs prediction. Most TISs start with AUG triplets in the RNA sequence. Additionally, six biological factors are influential in accurate TIS prediction, and the three agents are used in (97) as problem solvers. Prediction process contains five phases namely, solution generation, decision making, negotiation (mediator agent finds final prediction using Q-learning), execution, and feedback as it is shown in Figure 8 (97). The proposed method was evaluated on three datasets (vertebrates, Arabidopsis thaliana, and TIS+50) improving accuracy from 8% to 35%.

In genetics, the functioning units of DNA contain a group of genes controlled by the same promoter called an operon. The operon prediction is critical in regulating the networks in newly sequenced genomes (98). To tackle this problem, binary particle swarm optimization (BPSO) can be used to predict operons(99). The combination of BPSO with RL was investigated as well (100). The fitness function was formed based on the three main properties defining an operon and RL. These properties are intergenic distance, metabolic pathway, and gene length ratio. Assigning 1 to a pair of genes in one operon (OP) and 0 to others (NOP) at the end of the sequence yields a code representing the current prediction that can be used in BPSO algorithm. In the training phase, it uses the E.coli genome sequence dataset and three other datasets (B.subtilis, P.aeruginosa PA01, and S.aureus) in the evaluation phase, showing over 2 to 15% increase in the prediction accuracy.

CpG islands are another feature in genomics that can be predicted with machine learning techniques (101, 102). A CpG island is defined as a 200bp region of DNA containing a CpG dinucleotide range of more than 50% and an observed CpG versus expected CpG ratio greater or equal to 0.6 (GGF criteria) (102). Early studies mainly relied on sliding windows to identify CpG islands. However, their performance relies heavily on the window size. To address this issue, complement particle swarm optimization algorithm combined with RL (CPSORL) was proposed (103). To extend the length of found CpGs, the RL method combines CpGs provided that the distance between them is less than 200 bps. The reward function is defined based on empirical verification of CpGs. Five randomly selected contigs in chromosomes 21 and 22 of the human genome were used. Each contig is a set of overlapping segments of DNA representing a consensus region of DNA. CPSORL outperformed evaluated methods by finding larger islands (with higher CpG island methylation densities) in a shorter amount of time.

A ligand is an ion or molecule which forms a coordination complex by binding to a central metal atom or ion via donating pairs of electrons to them. Predicting the best ligand pose is a crucial step in protein-ligand interaction. The pose prediction was realized using Deep Deterministic Policy Gradient (DDPG) and it was used as the reward function. Thirteen viral protein-ligand complexes were selected from PDBbind and RCSB PDB. The proposed method reduces manual intervention and can also find the binding site(104).

There has been an attempt to find the most accurate architecture of convolutional neural networks (CNNs) for predicting regulatory features such as transcription factors, histone marks, and chromatin accessibility. The goal was achieved primarily by CNN (42, 105, 106) but preceding that, neural architecture search (NAS) algorithms were used which are very difficult with high computational load (107). To remedy the computational burden, RL-based method called AMBIENT was presented (44). In AMBIENT method, the optimal architecture for a given task is generated based on the current knowledge instead of training from scratch, reducing the computational demand.

### *C.2. Features assignment*

One of the essential steps in every genome project is the annotation phase which includes assigning biological functions to DNA and RNA sequences. This phase is divided into automatic (computer algorithms) and manual (biologists' knowledge) annotation. A Bio-Agent with three layers (interface, collaborative and physical) was introduced to support and simulate manual annotation (108) and one learning layer was added to the RL agent to choose the correct annotation from candidate solutions (109). Training was carried out using Paracoccidioides Brasiliensis and Paullinia Cupana datasets. The validation of the Bio agent suggestions was done based on reference genomes Caenorhabditis Elegans and Arabidopsis Thaliana, respectively. Reported results show about 6% increase in accuracy. The number of suggestions increased which as well was declared promising by biologists.

### *C.3. Dataset preparation*

Even the best learning methods without having access to sufficient amount of training data fail miserably. Therefore, gathering training data is crucial. However, depending on the learning problem properties, data collection may be expensive in terms of time and money. Utilizing learning methods such as RL has the potential to remedy data gathering costs which achieves human-like searching ability. For example, Wang et al. (110) proposed Everlasting Iatric Reader (Eir) method which relies on DQN to build a database containing reliable genetic associations between complex traits of humans and their genes. The reward function was set such that paper selection and final extraction accuracy were minimized. To keep track of agent state, bidirectional LSTM was employed.

As another example of RL utilization for dataset collection, Fan et al. (111) cast the cryo-EM data collection problem into an optimization task leading to proposal of cyroRL method. The DQN was used to plan the data collection. Although cyroRL method needs a pre-selection from a raw dataset, compared with previous methods, it is much less labor-intensive. Another paper used IRL and multi-fidelity Bayesian optimization (MFBO) to estimate the reward function that describes the biologist's actions during the intervention process and data collection (112). Testing this method on genomics, metagenomics datasets, and sets of random simulated problems demonstrates the reliability and scalability of the method.

### *C.4. Disease diagnosis and treatment*

Some of existing papers have relied on RL to tackle disease diagnosis and treatment based on genomic data. For example, actor-critic architecture was used to determine cancer types according to mutation patterns (113). The critic was implemented as a LSTM network and the actor was a SARSA agent. The samples were from patients with p53 mutants, used to determine the effect of signatures in mutations for the most likely cancer and its subtypes.

Given that MicroRNA (miRNA) is closely associated with complex diseases such as cancer, the possibility to predict human microRNA association with diseases based on Q-learning was investigated (114). The weighted predictions of three different models were aggregated to determine the final prediction. The method was tested on a benchmark dataset (HMDD database) leading to improved accuracy.

While deep learning has already been used to diagnose Alzheimer's disease (115), approaches based on DRL are still missing despite having great potential for disease diagnosis. Therefore, investigating the pros and cons of DRL on disease diagnosis is an excellent direction for future research.

RL methods are also used in drug discovery for mono and multi-genic diseases. The existing methods and their challenges like model and reward function definition, generalization, multi-objective optimization, and robustness have been reviewed (116). As an example of RL application, a generative de novo molecular design model was trained using policy gradient based on gene expression profile of the target cell or cancer site. The reward is provided by PaccMann module (117) which predicts anticancer drug sensitivity to generate molecules against cancer cell lines with low toxicity. The RL state space is the set of all possible molecular strings, and the action set consists of the canonical SMILES (Simplified Molecular-Input Line-Entry System) language symbols. The evaluation of the method on four different cancer types revealed that it can generate similar structures with known anticancer compounds.

Other than drug design, RL has been used to develop personalized treatments and recommend drug dosage for humans and other species. However, instead of genomic data, most of these studies tried to use biometric data of the subject as input due to better accessibility (118). As an example, a personalized drug ranking system called Proximal Policy Optimization Ranking (PPORank) (96) has been proposed. PPORank has been evaluated on two cancer cell line datasets namely Genomics of Drug Sensitivity in Cancer (GDSC) and Cancer Cell Line Encyclopedia (CCLE). The RL agent used in the drug ranking system is PPO (119). The reward function is the discounted cumulative gain (120) and the state space contains the current rank of candidate drugs and cell line data. Although the results were promising, more data efficiency is needed. While choosing the best drug for treatment is vital, minimizing the side effects of the disease treatment is just as important. In an attempt to control tumor progress and treatment side effects on lung tissues, DQN has been used to adapt radiation doses for cell lung cancer (CLC) patients (121). The data collection was done using clinical, genetic, and imaging radiomics data of 114 patients with radiotherapy experience. The dataset was destined for development of a radiotherapy artificial environment (RAE). The reward function is the uncomplicated cure probability (P+) which leads the agent toward a dosage similar to the clinical protocol. Inspired by the potential of RL-based disease diagnosis systems, Eckardt et al. (122) devoted a full survey article to the advancements in oncology based on RL and their corresponding challenges.

### *C.5. Features modeling*

RL also can be used in modeling some functions and mechanisms in cells. Protein synthesis is a core biological process occurring inside cells by ribosomes, guided by mRNA fragments. Translation elongation is an elusive critical step of protein synthesis. To understand the procedure, prediction, and modeling of ribosome distribution along the given mRNA sequence have been realized by a DRL-based method called Riboexp (123). This method outperforms the state-of-the-art approaches in predicting ribosome density by up to 5.9% on the datasets from three species and indicates a more informative sequence feature.

Plant breeders mainly use genomic selection to pick individuals for producing new generations of plants (124). This process involves selection, mating, and resource allocation (budget and time) to make sure that the picked individuals carry out their chosen traits to the next generation (89). The selection and mating tasks are optimized by look-ahead selection (LAS) (125). RL has proved useful in resource allocation of plant breeding programs (126). The resource allocation problem was cast into a Markov Decision Process and the action selection was done based on action-value function modeled as a random forest. The resources to be managed were number of performed crosses and number of progenies produced from each cross. An overview of articles, their concerning problem, and proposed solutions are summarized in Tables caption

**Table 1.** A comparison between the reviewed techniques in this study based on performance, network size and their main contribution.

| | Network type | Method | Performance (%) | Key points |
|---|---|---|---|---|
| **Control** | 7-gene WNT5A | Dynamic programming | 67 | One of the first studies that used RL. |
| | 10-gene Melanoma-related | Q-learning | 67 | Model-free and linear complexity |
| | 8067-gene Large scale Melanoma | Bach RL (FQI) | 80 | Large networks |
| | 7-gene Partially observable melanoma | LSFTDI | 90 | Partial observability |
| | 6-gene generalized repressilator | FQI | Tracking was done | Tracking control |
| | 2-gene genetic toggle switch | FQI | Regulation was made | Online and Offline training with no prior knowledge |
| | 7-gene melanoma and yeast | FQI SARSA | 53 | No need for time series data |
| | 7-gene melanoma | SGP-SARSA | 90 | Uncertain inputs and measurements, model-based |
| | 11-gene yeast cell cycle and 10-gene Metastatic Melanoma | BOAFQI-SARSA | 99 | Partial observability, used BOA |
| | -gene Melanoma and a p53-MDM2 negative feedback | BIRL+V_BKF+ Q-learning | 90 | Partial observability, unknown reward |
| | 7 and 25-gene melanoma | SQFT | 95 | Different feature vectors, more exploration |
| | 1000 to 7000-gene synthetic BNps | mSFQI | 87 | Used BOA, partial observability |
| **Modelling** | coupled oscillations in the cell cycle & cAMP receptor | - | 95 (accuracy) | Used ABGRN |

**Table 2.** Comparison of the performance of reviewed studies based on the task they involved. The "Key Point" column summarizes the main contribution of each study, and the "Performance" column reports the results as reported in the article, based on the defined reward function.

| | **Max length (bp)** | **Data** | **Method** | **Generality** | **Performance** | **Key points** |
|---|---|---|---|---|---|---|
| **Designing** | 8 | Biological sequences (protein, anti-microbial peptides) | Dyna-PPO | - | 0.7 | Offline model-based pre-training |
| | - | RNA sequences & secondary structures | **DQN and PPO** | Yes | 0.5 | RNA energy optimization |
| **Sequence Alignment** | 1093 | several DNA sequences (Hepatitis c virus, Papio Anubis , e.t.) | Q-learning | Yes | 0.45 - 0.95 | Low computational complexity |
| | 1093 | 8 sequences from mitochondrial human DNA | Actor-Critic | Yes | 0.96 | Fast convergence |
| | - | HEV, E.coli and mammalian DNA | **DQNalign** | Yes | 0.8 | Artificial dataset |
| | - | HEV DNA sequences | **DQNalign + x-drop** | - | 0.78 | Low computational complexity |
| **Assembly** | 8 | segments of E.coli DNA | Q-learning | - | - | High computational complexity |
| | 50 | 18 genome segments of E.coli | Q-learning | No | 0.2 | Inconvenient reward function |
| | 4000 | 23 genome segments of E.coli | GA+Q-learning | No | 0.8 | Better exploration, Pruning |
| **Rearrangement** | 10 | random permutations | **TD-lambda +DDQN** | No | Low | Pre-training |
| | 15 | random permutations | **WP+DDQN** | No | High | Memory fill |

**Table 3**.

### *C.6. Single-cell genomics and RL*

Single-cell genomics (ScG) data analysis has emerged as a transformative area of research, focusing on the individuality of cells and the diversity within complex biological systems (127). Single-cell sequencing spans all major modalities, including DNA, RNA, protein, and chromatin modification, providing a high-resolution view of cellular heterogeneity (85). ML techniques have been pivotal in processing these large and noisy datasets, transforming them into meaningful, low-dimensional representations suitable for downstream analysis (128). However, the application of RL to single-cell genomics remains limited due to several inherent challenges. A primary obstacle is the noisy and high-dimensional nature of single-cell data, which exacerbates issues such as sparse rewards and convergence instability in RL algorithms. Sparse rewards, where feedback is provided only at the end of an episode or upon achieving specific outcomes, can make it difficult for RL agents to learn efficiently, particularly in complex biological environments. Furthermore, single-cell data often lacks robust ground truth labels, making it challenging to define clear and biologically meaningful reward functions. This limitation hinders the development of RL models that can reliably adapt to the complexities of single-cell datasets.

Another critical barrier is the high computational cost associated with RL methods in single-cell genomics. Training RL agents often requires iterative interactions with the environment, which, in the case of single-cell data, involves large-scale simulations or experiments that are computationally intensive and time-consuming. While DRL has been employed to tackle high-dimensional problems in other domains, its applicability to single-cell genomics remains constrained by the need for extensive computational resources and optimized algorithms (129).

Interpretability is another significant challenge in applying RL to single-cell genomics. Understanding the logic behind RL decisions is crucial for biological problems, where insights must align with experimental validation and biological plausibility. The "black-box" nature of RL, particularly DRL, makes it difficult to discern why specific actions were taken, limiting its usability in critical tasks such as cell fate prediction or identifying regulatory interactions (86). Efforts to develop interpretable RL models or hybrid approaches combining RL with traditional ML methods may address this issue in the future.

Despite these challenges, some emerging studies indicate the potential of RL in single-cell genomics. For example, Single-cell Reinforcement Learning (scRL) has been proposed as a novel framework utilizing an actor-critic agent to analyze single-cell sequencing data and predict cell fate decisions (130).This approach addresses the critical challenge of early detection in developmental biology, allowing for forecasting and potentially perturbing cellular fate. By leveraging RL's adaptive learning capabilities, scRL demonstrates the feasibility of RL in scenarios where meaningful reward functions can be defined and computational constraints are manageable.

### *C.7. RL in combination with Large Language Models*

Reinforcement learning holds immense potential when combined with advanced methodologies like Large Language Models (LLMs), which are increasingly being applied across genomics and biomedical fields. Leveraging the strengths of both RL and LLMs can significantly enhance data analysis, interpretation, and decision-making processes. LLMs have already demonstrated substantial impact in molecular biosciences, healthcare, and neuroscience, with applications ranging from predicting DNA functionality and analyzing chromatin modifications to optimizing drug discovery pipelines (131). The integration of RL with LLMs offers additional advantages by enabling adaptive decision-making in tasks such as drug discovery and disease prediction, where models learn from interactions with complex biological datasets (132). For instance, in healthcare, reinforcement learning from human feedback (RLHF) can improve diagnostic accuracy and treatment recommendations, paving the way for more personalized and efficient care. However, integrating RL in genomics and biomedicine is not without challenges, including defining suitable learning objectives and generating high-quality datasets for training (133).

In genomics, RL can optimize the training of gene-language models to represent entire genomes by refining reward functions and enhancing learning strategies through feedback. This approach can lead to improved model performance across diverse genomic datasets (134). Additionally, RL can fine-tune genomic language models (GLMs) and DNA language models, automating complex tasks such as genomic sequence interpretation, gene-disease association prediction, and protein structure analysis. These advancements enhance predictive accuracy in personalized genomics while addressing challenges such as data scarcity and model interpretability (135). For example, recent research employing a three-track neural network for protein structure prediction demonstrated that RL can enhance model efficiency and prediction accuracy by incorporating adaptive learning strategies (136). Similarly, the Nucleotide Transformer, used for genomic sequencing, showcases how RL techniques can refine transformer-based model parameters, improving predictive capabilities (137). Lastly, ChatNT, a conversational agent for genomic tasks, highlights the use of RL to enhance user interaction and engagement, facilitating better communication and collaboration in biological research (138).

## III. Discussion

Reinforcement learning is a powerful technique for solving problems that involve decision-making under uncertainty. In recent years, there has been growing interest in applying RL to the field of genomics, with the goal of developing new methods for analyzing and interpreting large-scale genomic data. Various RL-based models have been proposed to analyze high-throughput sequencing data, model gene regulatory networks, and optimize drug discovery (14, 15). However, there are several challenges that need to be addressed in order to successfully apply RL in genomics.

A primary obstacle in applying reinforcement learning to genomics is the intricate and high-dimensional nature of genomic data, which can hinder the identification of meaningful patterns and relationships. Reinforcement learning algorithms typically rely on Markov Decision Processes (MDPs) to model the environment and the agent's interactions (139). However, in the genomics domain, the vast number of potential states and actions can render the accurate modeling of MDPs computationally challenging.

This can lead to poor performance of RL algorithms and high computational costs. Additionally, RL algorithms often rely on trial and error approach, which can be computationally expensive and time-consuming when dealing with large-scale genomic data. Additionally, this feature is a main drawback in real genomic applications where there is no place for miscalculations and can lead to disastrous outcomes, so it needs to be used with delicacy. Another challenge is the limited availability of labeled data for training RL algorithms (17). In genomics, there is often a lack of labeled data, such as functional annotations or drug response data, which can be used to train RL algorithms. This makes it difficult to learn accurate models of the environment and the agent's interactions with it. Checking the reviewed papers in section III reveals that learning problems in genomics domain are mostly based on discrete action spaces which is why SARSA, Q-learning, and its successor DQN are used several times in the reviewed papers (65). Another motivation for using the aforementioned algorithms is that they are model-free. This is advantageous because learning accurate environment models is usually challenging in real-world problems (28). However, model-free methods need more interaction with environment for appropriate learning which may be time consuming. A good compromise is settling down with hybrid methods that have model-free and model-based phases. This way, environment model is learned in the online model-free phase and exploited in the offline model-based phase.

To tackle these complexities, recent efforts have focused on improving the feature selection process in RL models. Techniques like auto encoders and feature selection algorithms are used to reduce dimensionality before training starts. These methods help manage computational demands and make the learning process more efficient, though maintaining accuracy is still a significant challenge. Enhanced feature selection is also supported by the use of transfer learning strategies, where pre-trained models on related genomic tasks are adapted to new, similar problems. This approach significantly reduces the need for large amounts of training data and computational resources.

The use of RL extends beyond genomic data analysis into the broader field of medical decision-making systems. RL's ability to manage complex decision processes, where outcomes are uncertain and decisions are made sequentially, makes it a promising tool for optimizing treatment strategies, patient management, and personalized medicine. In these systems, RL can optimize decisions in real time based on new data, adapting treatments to a patient's changing condition. For example, in managing chronic diseases like diabetes, RL can be used to continuously adjust insulin doses based on patient-specific responses. Additionally, RL has the potential to enhance clinical decision support systems (CDSS), which help physicians make timely decisions. By incorporating RL algorithms, CDSS can evolve from static tools into dynamic systems that learn from each interaction and improve their recommendations over time. Although RL methods are based on trial and error which is not a desirable feature in this field but its ability to learn and adapt can significantly improve patient outcomes by providing highly personalized recommendations that consider not only the patient's medical history and current state but also similar cases and broader medical knowledge.

Reviewed papers also reveal that, despite achieving promising results, classic methods like Q-learning may not be able to keep up with the growing complexity of genomics applications and more sophisticated approaches may be needed. This claim is justified by the fact that more advanced RL methods, such as PPO and custom actor-critic frameworks utilizing CNNs (82), are present among the reviewed papers. In addition, there is a need for robust and efficient RL algorithms that can handle large-scale data, account for uncertainty and missing data, and be able to generalize to new scenarios. Finally, the interpretability of the results obtained from RL in genomics is also a major challenge. The high dimensionality (61) and complexity of the data and models can make it difficult to understand the underlying biological mechanisms and interpret the results.

Importantly, reward functions are central to the success of RL agents, as they guide the learning process by defining objectives that align with the task at hand. In the context of genomics, designing reward functions involves domain-specific considerations to ensure the RL agent's actions are biologically meaningful and effective. For instance, in gene regulatory network control, reward functions are often defined based on desired gene expression states. Datta et al. (59) used a penalty-based reward function where the optimal state was defined as the downregulation of the *WNT5A* gene associated with melanoma, leading to a significant improvement in the probability of achieving this state. Similarly, in sequence alignment tasks, reward functions typically assign positive rewards for sequence matches and penalize mismatches or gaps. The DQNalign method (76)incorporated a reward schema where matches, mismatches, and indels (insertions or deletions) received varying weights to optimize alignment accuracy for sequences with low similarity.

Genome assembly tasks present additional challenges in reward function design due to the complex nature of sequencing. Bocicor et al. (85) developed a reward function based on the overlap scores of DNA fragments, optimizing the assembly of bacterial genomes like E. coli. This approach faced computational constraints, as the Q-table's dimensionality increased with the number of fragments, underscoring the importance of balancing reward specificity with computational feasibility. In disease modeling and drug discovery, reward functions often integrate multiple objectives, such as minimizing toxicity while maximizing efficacy. For example, PaccMannRL (33) used transcriptomic data to design anticancer molecules, with the reward function incorporating sensitivity predictions and structural viability.

Defining reward functions in genomics also comes with challenges. Sparse rewards, where feedback is only given upon reaching a specific state, can hinder efficient learning. Approaches like Hindsight Experience Replay (29) and adaptive reward shaping are employed to address this. Additionally, biological constraints, such as limited data availability and noisy measurements, necessitate robust reward functions that are resilient to inaccuracies. Best practices include iterative refinement of reward functions based on expert input and integrating domain knowledge to ensure relevance to biological objectives. Future research should focus on dynamic reward mechanisms that adapt based on intermediate states and outcomes, enhancing the applicability of RL in genomics.

Despite these challenges, the application of RL in genomics is rapidly growing. While there are still many challenges to overcome, the potential benefits of RL in genomics are clear, and it is likely that we will see continued progress in this field in the years to come.

# IV. Future Directions and Limitations

While applying RL and more generally ML to genomics problems is appealing, certain challenges need to be addressed for making further progress in the aforementioned fields. First, in ML literature, it is common practice to assume independence between dataset samples. This assumption is easily violated in genomics applications in which dependence between samples is fairly common and sometimes subtle which makes their recognition challenging (140). For example, some variables from sample $A_i$ might be correlated to other variables from sample $A_j$ which breaks the independence assumption. Putting sample $A_i$ in the training set and $A_j$ in the test set causes data leakage which could lead to learning unwanted patterns by the RL agent. Investigating approaches for avoiding data leakage is crucial in future work.

ML researchers always strive to come up with a model that captures the correlation between the input variables and the desired output. Failing to capture these correlations leads to models with poor performance. Given the complex relation between inputs and output of genomics applications, capturing the right correlations is challenging. Missing correlations may occur due to unmeasured or artefactual variables (140). The possibility to minimize the chance of missing important correlations between ML model inputs and outputs can be investigated in future studies.

Another naïve assumption in classical ML is that training and test sets have identical distributions. Yet again, distributional differences are quite common in genomics problems. Different epigenetic profiles and population genetic structures are among the sources of distribution mismatch (140). Dealing with distribution mismatch is necessary for the development of generic ML models. Domain generalization (DG) is an active field of research pursued by DL community to reduce the effect of distribution mismatch (141). However, DG is only part of the solution. Having solid theoretical and practical knowledge about ML approaches and the problem at hand is necessary for achieving reliable predictive models (142).

Reinforcement learning presents promising avenues for addressing a myriad of challenges in genomics. By devising reward functions that prioritize accuracy, precision, and recall, RL can optimize variant calling performance. Moreover, leveraging pre-trained models can enhance variant calling accuracy in specific domains. In drug discovery, RL can streamline virtual screening processes and even design novel drug molecules with desired properties. For protein structure prediction, RL can improve accuracy and efficiency, particularly for complex proteins. RL can also be applied to infer gene regulatory networks from gene expression data and develop models for disease diagnosis. In personalized medicine, RL can optimize treatment plans and identify potential new uses for existing drugs. Furthermore, RL can optimize CRISPR guide RNA design and predict the outcomes of genome editing experiments. In metagenomics, RL can analyze microbial communities and combat antibiotic resistance. By focusing on these specific areas, researchers can harness the power of reinforcement learning to advance our understanding of biological systems and address critical challenges in genomics.

Reinforcement learning holds great promise for genomics, but its application faces several challenges. One major limitation is the scarcity and quality of genomic data, hindering effective model training. Additionally, RL algorithms can be computationally intensive, especially when dealing with large datasets and complex reward functions. Designing appropriate reward functions can be subjective and challenging, as success in biological contexts is often complex to define. Furthermore, the black-box nature of RL models can make it difficult to interpret their decisions, hindering their adoption in critical applications.

To address these limitations, future research should focus on data augmentation and generation techniques to improve data availability and quality. Developing more efficient algorithms and utilizing specialized hardware can help reduce computational costs. Transfer learning and hierarchical RL can aid in designing effective reward functions, while model-agnostic explanations and causal inference can enhance interpretability. Integrating RL with other machine learning techniques can further improve performance and address limitations. By addressing these challenges and pursuing these future directions, reinforcement learning has the potential to make significant contributions to genomics research and applications.

DRL has better representation power compared to RL which makes it possible to tackle problems with higher dimensions. However, black box nature of DL deprives the user from gaining clear understanding of its working mechanism. Working on approaches to make DL behavior interpretable is an active field of research. Interpretable DL models are highly desired especially in medical applications that require thorough analysis (15). To analyze the decision process of DL models, tools like Captum (143) can be helpful.

# V. Conclusion

Reinforcement Learning has gained significant attention in recent years for its ability to overcome various problems in fields such as search, decision-making, robotics, and genomics. This paper reviews the applications of RL in different areas of genomics, including gene regulatory network control and modeling, sequence alignment, genome assembly and rearrangement, operon, TIS, and CpG island predictions. The use of gene expression data as a means of testing RL agents during development is also highlighted. One of the main challenges in gene expression data analysis is the sheer volume of raw data generated. Traditional techniques can struggle to extract meaningful patterns from such large datasets. However, the learning mechanism of RL agents, which involves the acquisition of (locally) optimal behavior with limited prior knowledge, makes it well suited to handle such data in genomics applications. While RL has demonstrated excellent performance in the control of gene regulatory networks, there are still open challenges in this area. The use of Deep Reinforcement Learning has the potential to overcome these limitations, as it can eliminate the restriction on network size and pave the way for control solutions in natural biological systems. In conclusion, RL approaches are proving to be effective in overcoming the limitations of traditional techniques in gene expression data analysis. Additionally, the potential of DRL to achieve even greater milestones in this field highlights the importance of further research and development in this area.

## CRediT author statement

HAR developed the concept. HAR, MK, KJ developed the method. MK, RA, FH, JW, MY, KJ, HAR, and FH wrote the manuscript. MK, RA, HAR, ID, RA, JG, YZ, MY revised the manuscript. MK, RA, and KJ analyzed the data and created all figures and tables. All authors read and approved of the final manuscript.

## Competing interest

The authors declare no conflict of interest.

## Acknowledgments

Analysis was made possible with computational resources provided by the UNSW BioMedical Machine Learning Laboratory (BML) Servers with funding from the UNSW Scientia Program Fellowship. This work was supported by the UNSW Scientia Program Fellowship; and the Australian Research Council Discovery Early Career Researcher Award (DECRA) under grant DE220101210 to HAR. This collaborative project was also supported by the Second Century Fund (C2F), Chulalongkorn University. Additionally, the project has support from the Ratchadaphiseksomphot Endowment Fund, Top-up Grant of C2F, Chulalongkorn University.

## Declaration of AI and AI-assisted technologies in the writing process

We would like to thank OpenAI's ChatGPT for its contribution in revising a portion of the manuscript (abstract and discussion sections). The assistance provided by the language model was invaluable in improving the quality of the written text

## ORCID

0000-0002-6506-2523 Mohsen Karami

0009-0005-5386-7673 Khadijeh Jahanian

0000-0001-8577-0271 Iman dehzangi

0000-0003-0898-5054 Roohallah Alizadehsani

0000-0002-8276-9774 Ahmadreza Argha

0000-0001-7069-1714 Juan M. Gorriz

0000-0002-4870-1493 Yudong Zhang

0000-0002-3165-7051 Jia Wang

0000-0002-8573-5297 Farshid Hajati

0000-0001-7345-5071 Min Yang

0000-0003-0145-9801 Thantrira Porntaveetus

0000-0002-2189-9153 Hamid Alinejad-Rokny

## Tables caption

**Table 1.** A comparison between the reviewed techniques in this study based on performance, network size and their main contribution.

| | Network type | Method | Performance (%)[1] | Key points |
|---|---|---|---|---|
| **Control** | 7-gene WNT5A | Dynamic programming | 67 | One of the first studies that used RL. |
| | 10-gene Melanoma-related | Q-learning | 67 | Model-free and linear complexity |
| | 8067-gene Large scale Melanoma | Bach RL (FQI) | 80 | Large networks |
| | 7-gene Partially observable melanoma | LSFTDI | 90 | Partial observability |
| | 6-gene generalized repressilator | FQI | Tracking was done | Tracking control |
| | 2-gene genetic toggle switch | FQI | Regulation was made | Online and Offline training with no prior knowledge |
| | 7-gene melanoma and yeast | FQI SARSA | 53 | No need for time series data |
| | 7-gene melanoma | SGP-SARSA | 90 | Uncertain inputs and measurements, model-based |

1 On reaching the desired network state

| | | | | |
|---|---|---|---|---|
| | 11-gene yeast cell cycle and 10-gene Metastatic Melanoma | BOAFQI-SARSA | 99 | Partial observability, used BOA |
| | -gene Melanoma and a p53-MDM2 negative feedback | BIRL+V_BKF+ Q-learning | 90 | Partial observability, unknown reward |
| | 7 and 25-gene melanoma | SQFT | 95 | Different feature vectors, more exploration |
| | 1000 to 7000-gene synthetic BNps | mSFQI | 87 | Used BOA, partial observability |
| **Modelling** | coupled oscillations in the cell cycle & cAMP receptor | - | 95 (accuracy) | Used ABGRN |

**Table 2.** Comparison of the performance of reviewed studies based on the task they involved. The "Key Point" column summarizes the main contribution of each study, and the "Performance" column reports the results as reported in the article, based on the defined reward function.

| | **Max length (bp)** | **Data** | **Method** | **Generality** | **Performance**[2] | **Key points** |
|---|---|---|---|---|---|---|
| **Designing** | 8 | Biological sequences (protein, anti-microbial peptides) | Dyna-PPO | - | 0.7 | Offline model-based pre-training |
| | - | RNA sequences & secondary structures | **DQN and PPO** | Yes | 0.5 | RNA energy optimization |
| **Sequence Alignment** | 1093 | several DNA sequences (Hepatitis c virus, Papio Anubis , e.t.) | Q-learning | Yes | 0.45 - 0.95 | Low computational complexity |
| | 1093 | 8 sequences from mitochondrial human DNA | Actor-Critic | Yes | 0.96 | Fast convergence |
| | - | HEV, E.coli and mammalian DNA | **DQNalign** | Yes | 0.8 | Artificial dataset |
| | - | HEV DNA sequences | **DQNalign + x-drop** | - | 0.78 | Low computational complexity |
| **Assembly** | 8 | segments of E.coli DNA | Q-learning | - | - | High computational complexity |
| | 50 | 18 genome segments of E.coli | Q-learning | No | 0.2 | Inconvenient reward function |
| | 4000 | 23 genome segments of E.coli | GA+Q-learning | No | 0.8 | Better exploration, Pruning |
| **Rearrangement** | 10 | random permutations | **TD-lambda +DDQN** | No | Low | Pre-training |
| | 15 | random permutations | **WP+DDQN** | No | High | Memory fill |

**Table 3.** An overview of studies in new application fields and their proposed methods. This table also lists the datasets and cost functions used.

2 On reaching desirable sequences

| | Problem | Method | Dataset | Reward (Cost function) |
|---|---|---|---|---|
| Predicting and Assigning | predicting TISs | MAS-TIS | mRNA (vertebrates, Arabidopsis thaliana, and TIS+50) | Expert decision |
| | Operon prediction | BPSO+RL | E.coli, Busuttil's, P.aeruginosa PA01, and S.aureus genome sequences | intergenic distance, metabolic pathway, and gene length ratio |
| | Identify CpG islands | CPSORL | five randomly selected contigs in chromosomes 21 and 22 of the human genome | Experimentally verified CpGs |
| | Optimize ligand pose | **DDPG** | 13 viral protein-ligand complexes from PDBbind and RCSB PDB | pose prediction function |
| | automatic annotation | MAS+RL | Paracoccidioides Brasiliensis and Paullinia Cupana | Expert decision |
| Dataset | building database | **DQN (Eir)** | Websites | paper selection, final extraction accuracy, number of queries |
| | More informative images | cryoRL | - | pre-selected images |
| | data collecting | IRL+ MFBO | genomics, metagenomics datasets | Based on biologist action |
| Diagnosing and Treatment | Cancer diagnosis | **Actor-Critic** | Samples from patients with p53 mutants and HMDD | Mutation Patterns |
| | microRNA-disease prediction | Q-learning | HMDD | Expert decision |
| | Personalize drug discovery | Policy Gradient | GEP (4 Cancer types) | PaccMann module (117) |
| | Drug recommendation | **PPO-Rank (DRL)** | Genomics of Drug Sensitivity in Cancer (GDSC) and Cancer Cell Line Encyclopedia (CCLE) | Discounted Cumulative Gain (DCG) |
| | Adapt radiotherapy dosage | **DQN** | clinical, genetic and imaging radiomics data of 114 patients | uncomplicated cure probability (P+) |
| Modeling | Ribosome density modeling | Riboexp | mRNA from 3 species | Experimental data |
| | Genomic selection | LAS+RL (Q-Learning) | 369 genotypes of maize inbred lines | genomic estimated breeding values (GEBVs) |

# Figures legend

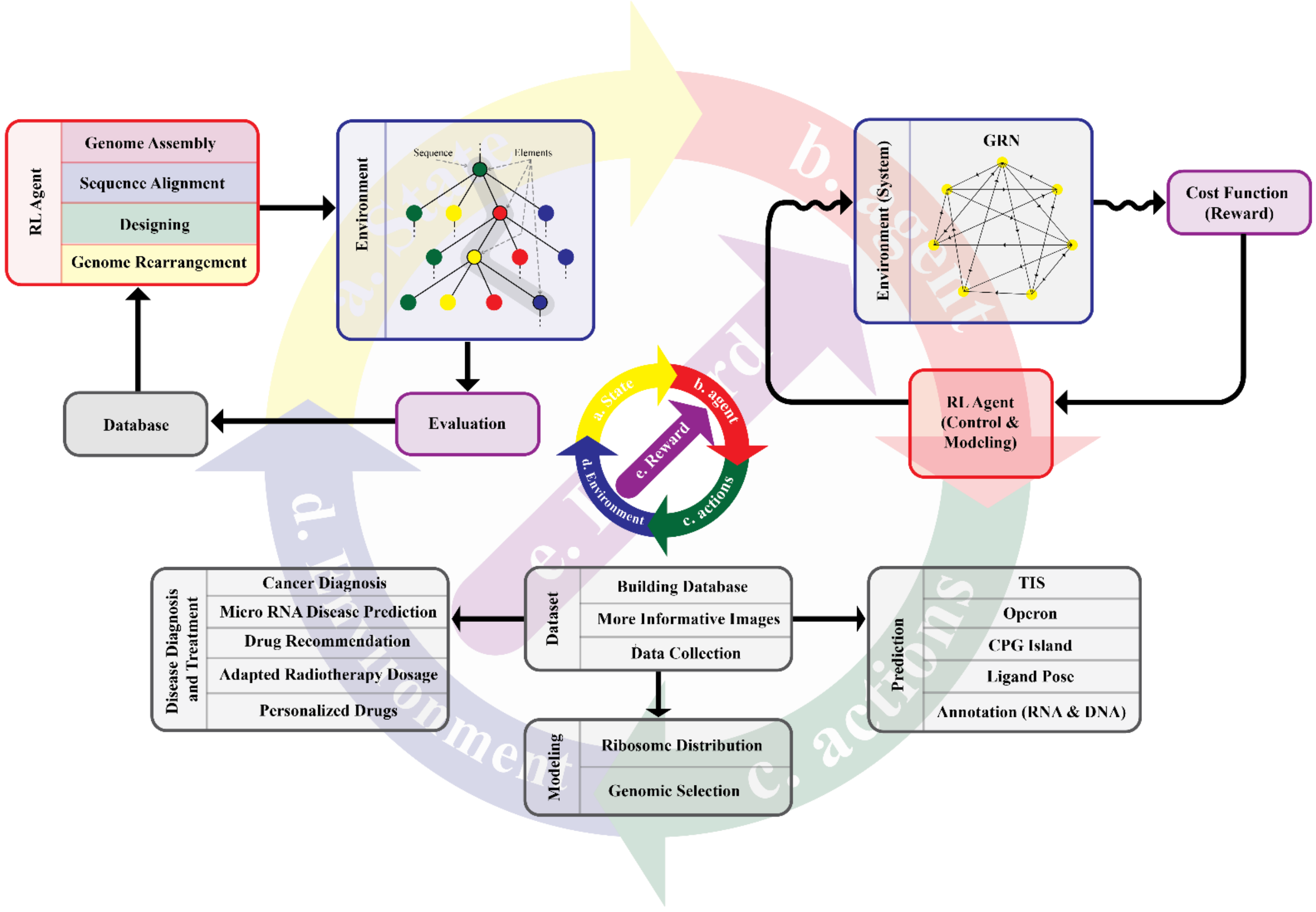

**Figure 1.** The Role of Reinforcement Learning in Genomics: RL has primarily been employed for investigating gene regulatory networks. The second most widespread application of RL is in the field of sequencing, such as genome assembly and alignment. Despite being less frequently utilized for tasks such as modeling and prediction, RL has demonstrated superior performance compared to alternative methods.

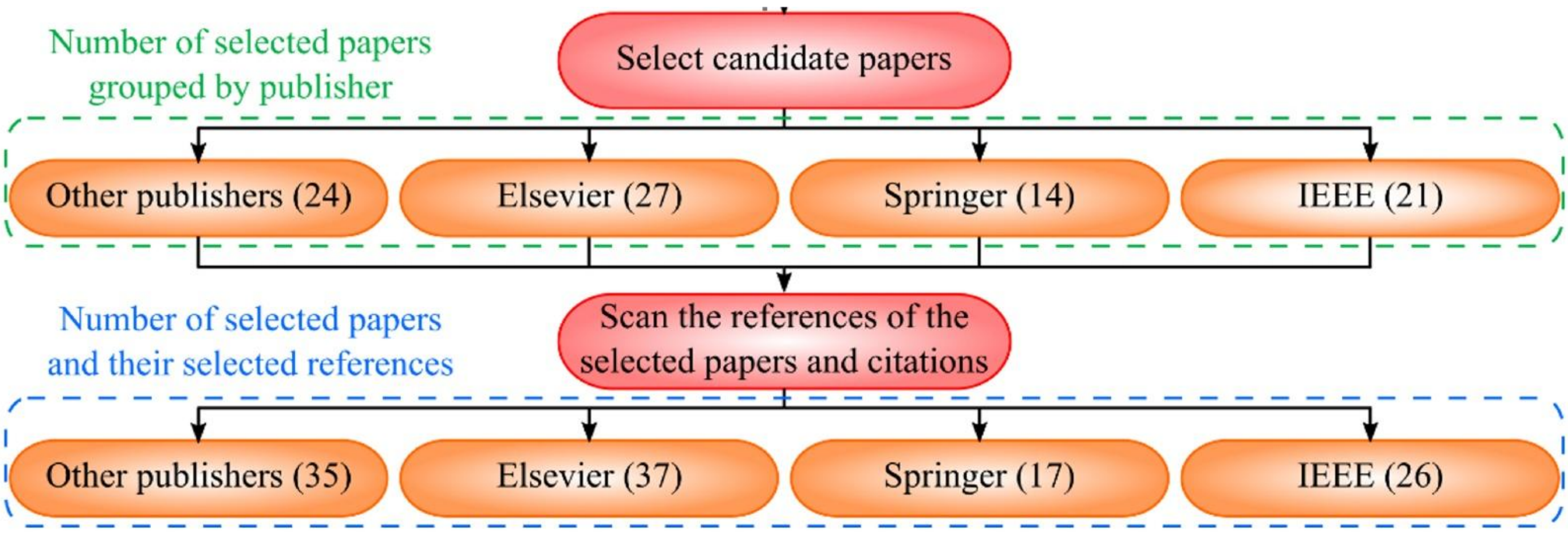

**Figure 2.** The flowchart of gathered papers in this survey.

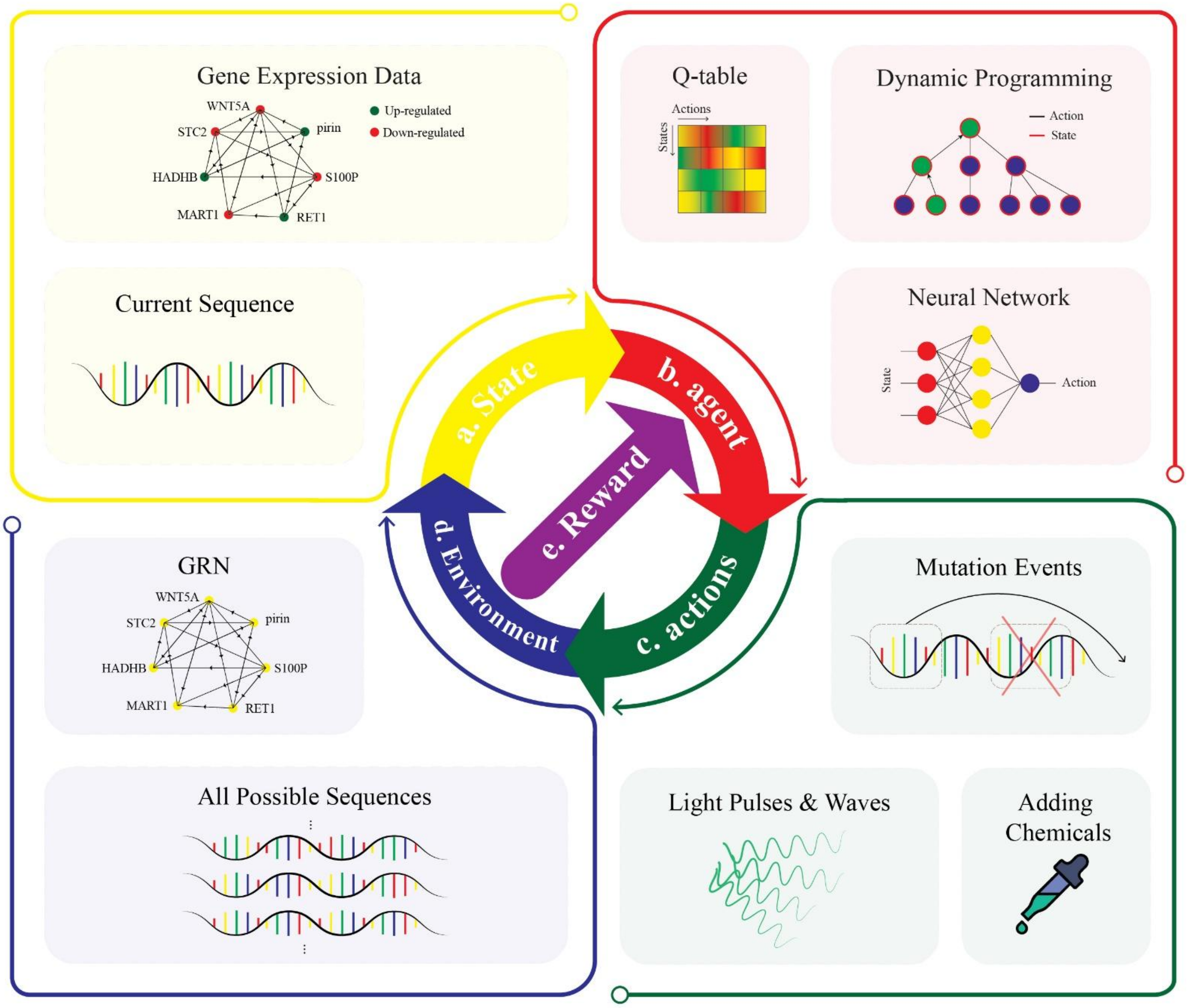

**Figure 3.** This figure illustrates the cycle of interaction between the reinforcement learning agent and the environment in genomics problems. The cycle consists of the following steps:(a) The environment state, such as gene expression data or DNA sequence, is transmitted to the agent. (b) The agent performs action selection. (c) The selected action is communicated to the environment. (d)

The action is executed in the environment, and a new state is determined. (e) The environment calculates the reward and sends it back to the agent. The cycle then repeats, with the agent using the new state to perform action selection and the environment responding with a new state and reward, until a stopping criterion is met.

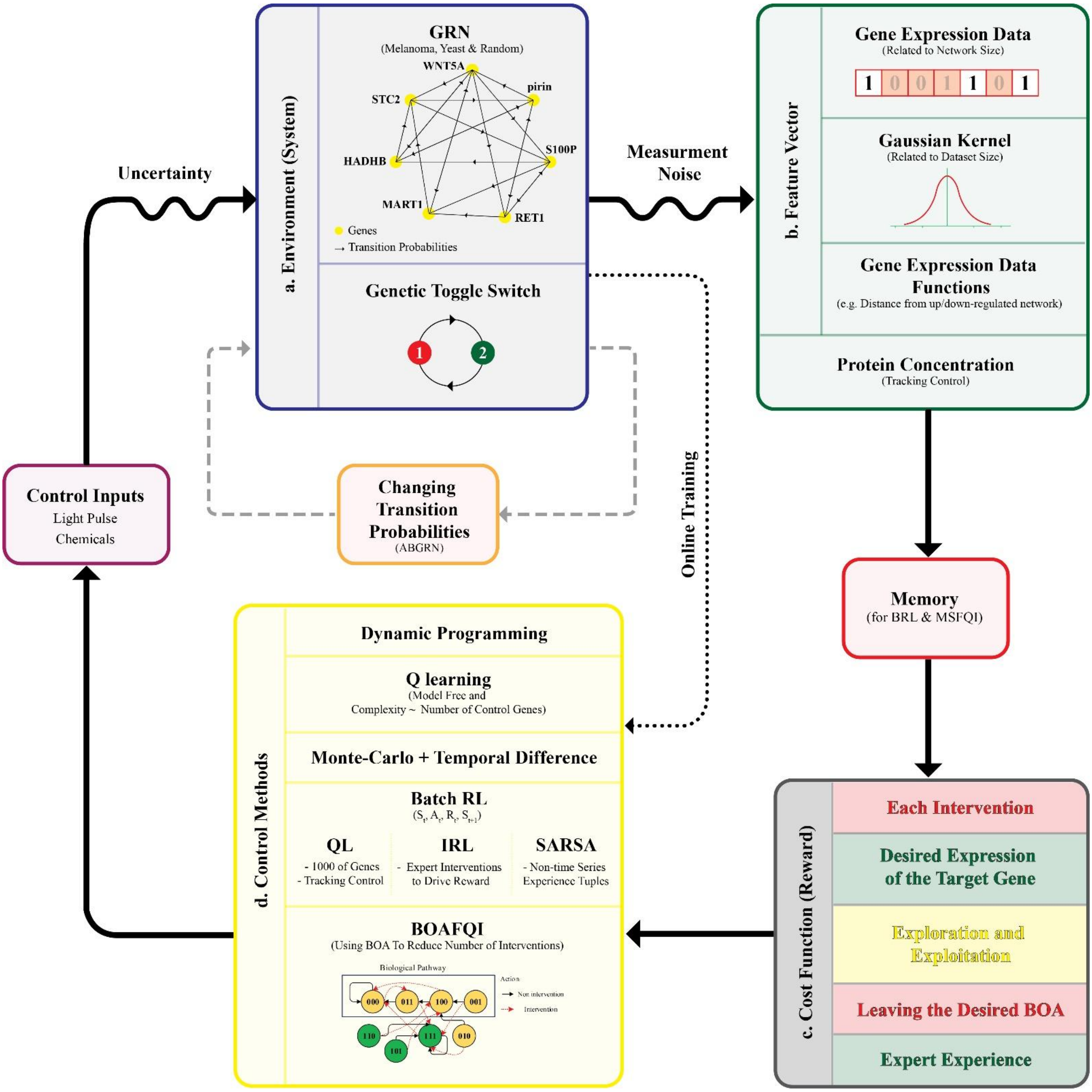

**Figure 4.** RL-based Control of Gene Regulatory Networks: (a) The GRN is treated as the RL environment, (b) The GRN provides the current state as a feature vector, (c) A cost function is used to calculate the reward for the current state, (d) The RL agent is trained to find the optimal action (control input) that will reach the desired state with the minimum number of interventions, for example, using Bayesian Optimization Algorithms (BOAs). RL is also utilized to determine the optimal network configuration (inner loop).

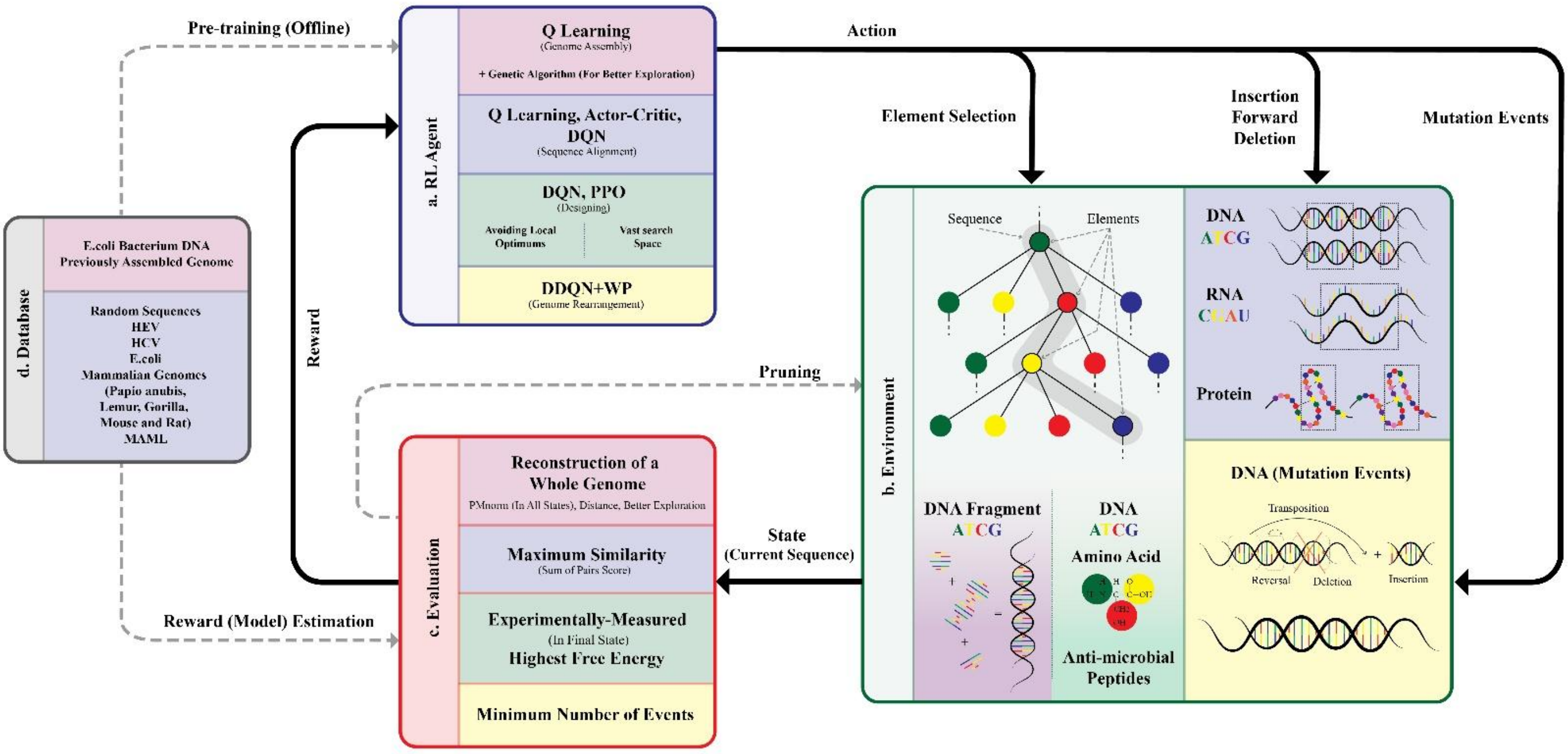

**Figure 5.** Interaction and training loop of RL in sequencing and rearrangement: (a) The RL agent selects actions. (b) The selected actions are executed in the environment, whose state is represented by a sequence, such as DNA or molecules. (c) The cost function provides the agent with reward values, based on which the agent optimizes the structure. (d) Available datasets are used for pre-training and reward estimation.

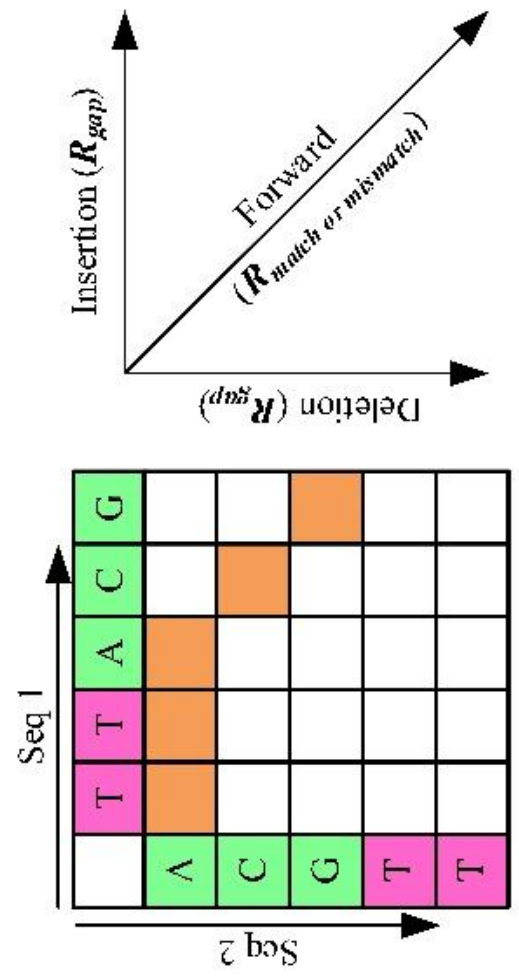

**Figure 6.** Local best path selection method.

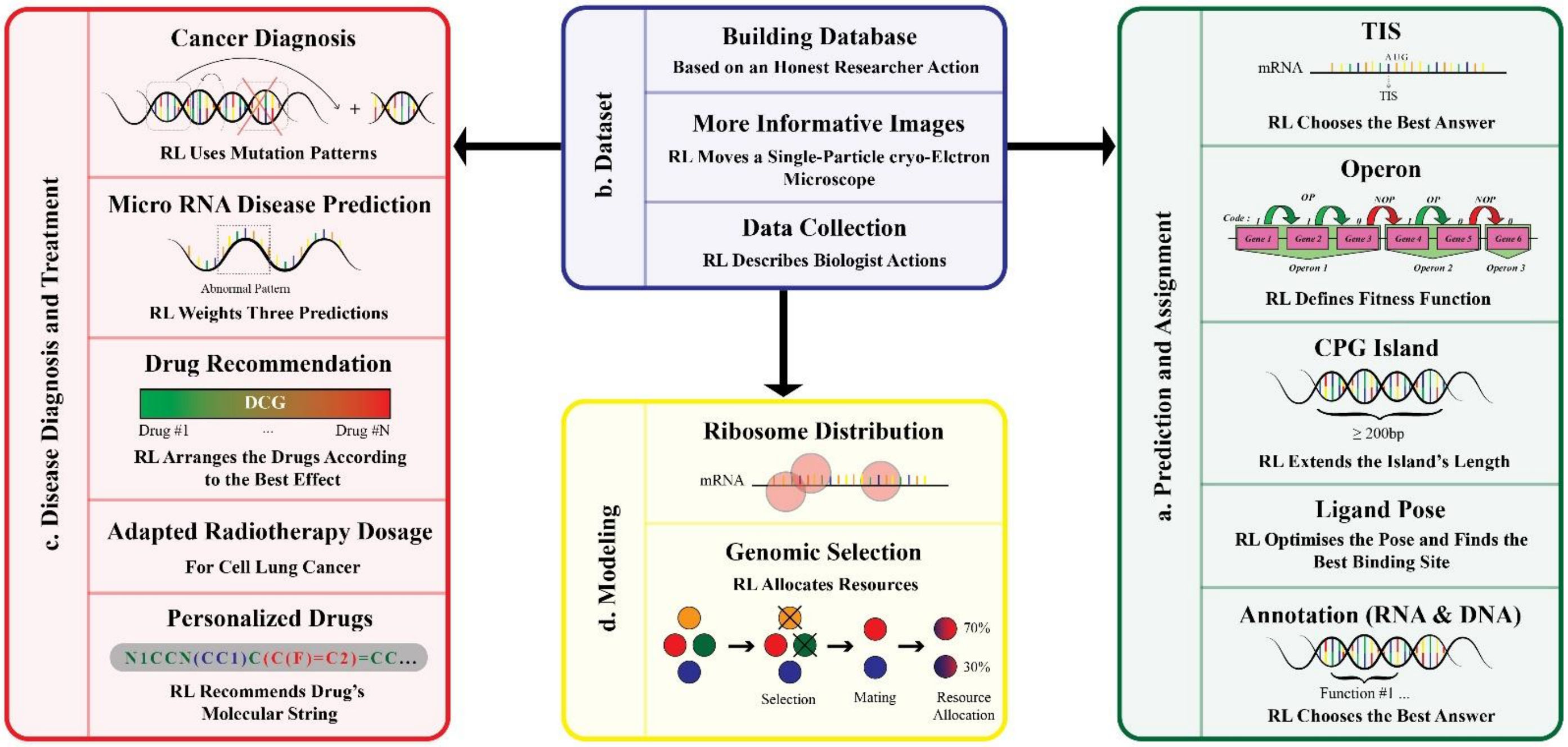

**Figure 7.** Application of Reinforcement Learning in other Domains: (a) Predictive Modeling and Task Assignment, (b) Data Preprocessing, (c) Disease Diagnosis and Treatment, and (d) Modeling. Reinforcement learning is capable of effectively addressing the challenging constraints in these four problem categories, demonstrating its broad learning potential.

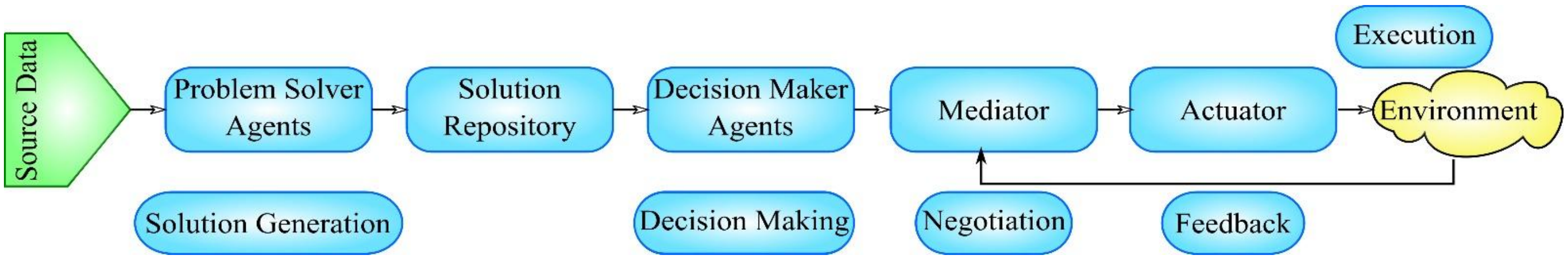

**Figure 8.** General Multi-Agent System-Tissue Information System Architecture. RL is used in the mediator component to aid biologists in selecting the optimal solution. This particular study employs Q-Learning as the RL algorithm.